%% file: paper.tex
\documentclass[british, 12pt, a4paper]{article}
\usepackage[left=0.8in,top=1in,right=0.8in,bottom=1in,nohead,paperwidth=8.5in, paperheight=11in]{geometry} %1 inch margins, 8 1/2 x 11-inch pages: standard choices for most journals
\usepackage[utf8]{inputenc}
\usepackage{fnpct}
\usepackage[dvipsnames]{xcolor} 
\usepackage[colorlinks=true,linkcolor=RoyalBlue, citecolor=RoyalBlue, urlcolor=blue]{hyperref}
\usepackage{geometry}
\usepackage{setspace}
\usepackage{rotating}
\usepackage{booktabs} 
\usepackage{adjustbox}  
\usepackage{graphicx,epsfig, color, subcaption}
\usepackage{xcolor}
\usepackage{threeparttable}
\usepackage[capposition=top]{floatrow}
\usepackage{tablefootnote}
\usepackage{bm}
\usepackage[affil-it]{authblk}
\usepackage{natbib}
\usepackage{array}
\usepackage{makecell}
\usepackage{float}
\usepackage{longtable}
\usepackage{dsfont}
\usepackage{amsfonts}
\usepackage{amssymb}
\usepackage{etoolbox}
\usepackage[per-mode=symbol]{siunitx}
\usepackage{adjustbox}
\usepackage{pdflscape}
\usepackage{caption}
\usepackage{subcaption}
\usepackage{xcolor}
\usepackage{lscape}
\usepackage{bbm}

\usepackage{amsmath,amssymb,amsthm,amsfonts,amsbsy}
\usepackage{mathrsfs}
\usepackage{libertine} % euro symbol
\usepackage{bbm} % indicator variable

\def \X {\mathbf{X}} 
\def\i {\mathds{1}}

\def\E {\mathbb{E}}

\def\F {\mathbb{F}}
\def\P {\mathbb{P}}
\def\G {\mathcal{G}}

\newcommand{\yit}{Y_{it}}
\newcommand{\yitau}{Y_{i\tau}}
\newcommand{\0}{\mathbf{0}}

\newcommand{\Gig}{G_{ig}}

\newcommand{\dmini}{\mathbf{d}_{-i,t}}
\newcommand{\Dmini}{\mathbf{D}_{-i,t}}
\newcommand{\hhi}{\mathbf{h}_{it}}
\newcommand{\hi}{h_{it}}

\newcommand{\Si}{S_{i}}

\newcommand{\indep}{\perp \!\!\! \perp}

\theoremstyle{plain}
\newtheorem{assumption}{Assumption}

\numberwithin{intassumption}{assumption}

\newtheorem{theorem}{Theorem}

\vspace{-1cm}  
\title{Environmental Policy and Firm Performance in Europe:\\ A Difference-in-Differences Approach with Spillovers}

\date{December 2025}
\bigskip 
\author{Andrea Ciaccio$^a$ $^b$, Francesco Moscone$^b$ $^c$, Elisa Tosetti$^a$\footnote{\footnotesize {\tt e-mail}: andrea.ciaccio.1@unipd.it,  francesco.moscone@unive.it,   elisa.tosetti@unipd.it. We thank the participants to the workshop ``From Consumption to Productivity: The Economic and Social Effects of Modern Regulation" held in Venice on the 19th May 2025, and the Ca' Foscari internal departmental seminar for their comments and advice. This article was supported by the PRIN project n. P2022BNNEY entitled "Board Gender Diversity and Resilience of European Firms'' funded by the MUR.}\\
\small{\textit{$^{a}$University of Padua, Department of Economics and Management, Italy}}\\
\small{\textit{$^{b}$Ca' Foscari University of Venice, Department of Economics, Italy}}\\
\small{\textit{$^{c}$Brunel University London, United Kingdom}}\\}

\begin{document}

\maketitle

\begin{center}
\mbox{}\\
\begin{minipage}{17cm}
\noindent\linespread{1}\selectfont    
\begin{abstract}
{\small
In this paper we investigate the causal impact of the European Union Emissions Trading System, a cap-and-trade scheme limiting greenhouse gas emissions of firms, on their environmental performance. Although previous studies have focused primarily on the effect of the emission cap imposed by the policy, we argue that the trading mechanism creates complex interdependencies among firms that can change the policy's intended effects. We develop a novel Difference-in-Differences approach that disentangles the direct causal effects of the scheme on regulated firms from the indirect spillover effects arising from trading among firms. By incorporating potential interference between treated units, our methodology allows a more comprehensive assessment of the policy's overall effectiveness. Monte Carlo simulations show that our proposed estimators perform well in finite samples, confirming the reliability of our approach. To assess the direct and indirect effects of the scheme, we construct a novel database on emissions of European industrial sites by matching information on treated plants from the European Commission's Community Independent Transaction Log with emission data from the European Pollutant Release and Transfer Register for the years from 2001 to 2017. We find that the scheme reduced emissions only for non-trading plants, but such reduction is entirely offset when accounting for spillovers from trading plants, thus suggesting that the trading mechanism neutralizes the environmental benefits of the policy. Our findings have important implications for the design of future environmental policies and the ongoing evaluation of cap and trade policies.
\newline

}

\end{abstract}
\end{minipage}
\end{center}

\section{Introduction}

The rapid escalation of environmental problems in the recent decades has highlighted the urgent need for innovations that can mitigate the environmental impact of economic activity. One of the most important policy responses to address emission reductions is the European Emissions Trading Scheme (EU ETS). The EU ETS is the largest international market for trading greenhouse gases, covering approximately 19,600 installations\footnote{An installation is a stationary technical unit where one or more of the activities are carried out, and any other directly associated activities which have a technical connection with those activities and which could have an effect on emissions and pollution. In this paper we will use the terms installation, plant and production unit as synonyms. See Article 15 of Council Directive 96/61/EC concerning Integrated Pollution Prevention and Control.} from over 11,000 firms in 31 European countries. Introduced in 2005, the system sets an annual cap on total greenhouse gas emissions on plants operating in sectors that are intensive in CO\textsubscript{2} emissions, such as power generation and the heavy industry. Within this cap, plants receive or purchase emission allowances, which they can trade with one another. This trading mechanism is intended to introduce flexibility in compliance and promote cost containment. The EU ETS was expected to reduce carbon emissions and pollution, as well as boost economic growth through the development of new low-carbon technologies \citep[see, for instance,][]{europa2024}. However, since its introduction, concerns have been raised about its effectiveness in reducing emissions, as well as its potential impact on the competitiveness of regulated businesses, potentially affecting economic growth. In response, a growing literature estimates the causal impact of the EU ETS on emissions, economic performance, and innovation by comparing regulated and unregulated firms before and after its implementation, often within specific countries (\cite{calel2020}, \cite{Dechezlepretre2023} and \cite{Colmeretal2024}; see Section \ref{empirical} for further details). While this literature primarily focuses on the effects of the emissions cap on firm- or facility-level outcomes, it has largely overlooked the additional impacts generated by the trading mechanism embedded in the policy. The EU ETS not only imposes an upper limit on total emissions but also creates a market for emission allowances, enabling firms to buy and sell emission allowances and introducing complex interdependencies. One important consequence of this trading mechanism is that the actual emissions released by each facility often diverge from what would have been emitted without trading. In addition, the overall impact of the policy is not solely determined by the cap itself, but also by how firms interact within the allowance market. In some cases, trading may attenuate the intended effects of the cap by allowing less efficient or more polluting plants to continue operating through the purchase of permits, potentially reducing the environmental benefits of the policy.\newline 

This paper aims to fill this gap by examining both the direct impact of the EU ETS emissions cap on plants' environmental outcomes and the indirect , and possibly offsetting, effect introduced by the trading mechanism.
To this end, we propose a Difference-in-Differences (DiD) approach that consistently estimates the direct average treatment effect of the policy on regulated firms (ATT), while allowing for spillover effects arising from interactions induced by allowance trading among \textit{treated} firms. Standard DiD frameworks typically rely on the Stable Unit Treatment Value Assumption (SUTVA) \citep{angrist1996identification}, which rules out interference by requiring potential outcomes to depend solely on own treatment status.\footnote{Hereafter, we use the terms ``spillover'' and ``interference'' interchangeably.} \newline 

A recent strand of literature develops identification, estimation, and inference procedures for causal parameters under interference. \cite{VazquezBare2023} proposes an estimator for average direct and spillover effects in the context of randomised controlled trial, where spillovers take place within non-overlapping groups of treatment and control units. \cite{Butts2021} 
considers a DiD approach for estimating the ATT when the policy generates local spillovers from treated to nearby untreated units, and further extends this method to staggered adoption following \cite{Borusyak2021}. Building on \cite{Wooldridge2021}, \cite{Fiorini2024} present a DiD method for settings without pre-treatment spillovers, where, after the implementation of the policy, a fraction of untreated units is exposed to spillovers from treated units. Unlike \cite{Butts2021}, their approach does not require spillovers to be local. Finally, \cite{xu2025} develops a DiD method that estimates both direct and indirect effects from a finite-sample population perspective. The approaches reviewed above, however, do not allow to carry causal inference in settings with non-local spillovers among treated units that arise after treatment adoption.
Consistent with our empirical problem, in this paper we propose a strategy for identification and estimation of the direct and indirect impact of the policy in the presence of spillovers among treated units, and extend the \textit{chained DiD} estimator by \cite{Bellegoetal2025} to explicitly incorporate spillover effects. Relative to previous studies on estimation of causal parameters under interference, our approach is more general and broadly applicable, as we consider a general setting where treatment effects are heterogeneous, treatment is introduced across multiple periods and groups with staggered timing, and the panel is possibly unbalanced. Allowing for an unbalanced panel is crucial in our empirical application, since--except for a small subset of firms--we do not observe firms throughout the entire sample period. This reflects the fact that firms may close, merge, or be acquired. Further, treated firms may be exposed to spillovers in some years but not in others; since our approach estimates the average effect of the policy using the subsample of units unaffected by spillovers, this subsample is necessarily unbalanced. 
In a set of Monte Carlo experiments, we show that our estimators perform well even for small samples, and that neglecting spillovers leads to severe bias, although such bias decreases as the fraction of units subject to spillovers reduces.\newline 

To assess \textit{both} the direct and indirect impact of the EU ETS in Europe, in the second part of the paper we construct a novel database at the facility level\footnote{A facility is one or more installations on the same site, operated by the same natural or legal person.} by matching information on treated facilities from the European Commission's Community Independent Transaction Log (CITL) with emission data from the European Pollutant Release and Transfer Register (E-PRTR). Our final data set contains information on $7,346$ unique facilities across 31 countries over the year from 2001 to 2017, for a total of $43,570$ facility-year observations. Relative to previous empirical studies, our analysis introduces several novel contributions. First, our data set includes facilities from across the entire European Union, providing comprehensive coverage of the regulated industrial sector. Second, the analysis is conducted at the facility level, whereas most existing studies rely on firm-level data. Examining effects at the facility level is crucial, as firms often operate multiple facilities that differ in production technology, abatement costs, and exposure to regulatory constraints. Third, we disentangle the effect of the emission cap from that of the trading mechanism, allowing us to isolate the distinct channels through which the EU ETS influences emission outcomes. Fourth, in contrast to previous studies, we explicitly account for both the staggered implementation of the policy and the unbalanced structure of the panel. Neglecting the staggered timing of treatment may result in biased estimates of the average causal effect in the presence of treatment effect heterogeneity \citep{Goodmanbacon2021}.\newline

We find that the EU ETS reduces emissions by approximately 26-29\% among treated facilities that do not engage in permit trading (direct effect). However, when spillovers from facilities actively engaged in allowance trading are taken into account, the effect substantially reduces and becomes statistically insignificant. These results are robust to conditioning on observable characteristics and to a range of robustness checks. A heterogeneity analysis shows that the policy effect is primarily driven by large emitters, possibly because of their greater capacity to adjust their production processes in response to regulation. Further, the policy effect is stronger in countries where environmental policies were less stringent before the reform. This suggests that nations starting from a weaker regulatory baseline experienced greater marginal improvements in response to the new policy. Our findings have important implications for the design of future environmental policies and the ongoing evaluation of the EU ETS as a tool to achieve climate objectives while maintaining economic competitiveness, providing evidence on how trading systems can generate unintended consequences through firm interactions. The approach proposed in this paper could be useful for evaluating other, possibly non-environmental, cap and trade systems in the world, or more in general, policy interventions that generate spillovers among units.\newline 

The remainder of the paper is organised as follows. Section~\ref{methods} introduces our Difference-in-Differences estimator with spillovers. Section~\ref{MonteCarlo} reports results from a Monte Carlo study assessing the finite-sample properties of the proposed estimator. Section~\ref{empirical} describes the EU ETS scheme and provides descriptive statistics. Section~\ref{results} presents the estimation results and a discussion of potential mechanisms behind the findings. Finally, Section~\ref{conclusions} concludes.

%----------------------------------------------------------------------------------------
%	THEORETICAL PART/ EMP. STRATEGY
%----------------------------------------------------------------------------------------
\section{Empirical strategy: DiD with spillovers}\label{methods}

In this section, we propose a general DiD framework to evaluate policy effects in the presence of spillovers among treated units, establish identification conditions, and present an estimation procedure applicable to both balanced and unbalanced panels, including rotating panels and arbitrary missing-data patterns.

\subsection{Setup}
Consider the case of $T>2$ time periods, indexed by $t = 1, \dots, T$, and let $i = 1, \dots, n$ index cross-sectional units. Define a binary treatment indicator $D_{it}$ equal to $1$ if unit $i$ is treated in period $t$, and $0$ otherwise. We impose the following assumption on the treatment assignment process 

\begin{assumption}[\textit{Irreversibility of Treatment}]\label{as_irreversibility}
$D_{i1}=0$. For all $t=2,\dots, T$, $D_{it-1}=1$ implies $D_{it}=1$.
 \end{assumption}

For ease of presentation, we will present identification results focusing on the setting with a never-treated comparison group, although they can be easily generalised to the case where not-yet-treated units serve as the comparison group. Let $G_i$ denote the first period in which unit $i$ receives the treatment, with $G_i = \infty$ if unit $i$ is never treated in any period, $G_{ig}$ be a binary indicator equal to $1$ if unit $i$ is first treated in period $g$ (i.e., $G_{ig}=\i\{G_i=g\}$), and $C_i$ be equal to $1$ if unit $i$ is never treated (i.e., $C_i=\i\{G_i=\infty\}=1-D_{it}$). Let $\bar g= \max_{i=1,\dots, n} G_i$ denote the last treatment adoption period in the sample, and $\G = \text{supp}(G)\backslash {\bar g}\subseteq\{2,3, \dots, T\}$ be the support of $G$ excluding $ {\bar g}$. If a unit is still untreated by time $s$, i.e., part of the not-yet-treated group, then $(1-D_{is})(1-G_{ig})=1$.
To model potential violations of SUTVA, we introduce an `exposure mapping', $h_{it}(\cdot)$ that summarises how the treatment status of other units affects the potential outcomes of unit $i$ in period $t$. To this end, let $\mathbf{D}_t = \left(D_{1t}, \dots, D_{nt} \right)'$ denote the vector of treatment assignments for all units at time $t$, with realization $\mathbf{d}_t\in \mathcal{D}\subseteq\{0,1\}^{n}$, and $\Dmini$ be the $(n - 1)$-dimensional vector of treatment indicators for all units other than $i$, with realization denoted by $\dmini$.  The exposure mapping is defined as a function $\hi:\mathcal{D}\xrightarrow{}\mathcal{H}$ that maps $\dmini$ into some value $\hi(\dmini)$ of the same or lower dimension. For $\hi(\dmini) = \hhi$, we refer to the tuple $(d_{it}, \hhi)$ as the effective treatment assignment of unit $i$, an element in the set $\{0,1\} \times \mathcal{H}$.
Then, for $g \in \mathcal{G}$, let $\yit(g, \hhi)$ denote the potential outcome for unit $i$ at time $t$ had the policy been introduced for unit $i$ in period $g$, and all other units treated according to $\hi(\dmini) = \hhi$. 
Hereafter, $\mathbf{0}$ will denote $(n - 1)$-dimensional vector of zeros. If $\hi(\dmini) = \mathbf{0}$ represents no exposure to spillovers, then the untreated potential outcome without interference is given by $\yit(0, \mathbf{0})$, whereas the treated potential outcome for units first treated in $g$ without interference is denoted by $\yit(g, \mathbf{0})$. The observed outcome for a generic unit $i$ at time $t$ is then given by:

\begin{equation}\label{observed_y}
    \begin{aligned}
            \yit = & \yit\left(0,\0\right)+  \sum_{g\in \G} \left(\yit\left(g,\0\right)-\yit\left(0,\0\right)\right)\cdot G_{ig} \cdot \i\{\hi(\dmini)=\0 \}\\
             & +  \sum_{g\in \G}\sum_{\hhi \in \mathcal{H}}\left(\yit\left(g,\hhi \right)-\yit\left(0,\0\right)\right)\cdot G_{ig}\cdot \i\{\hi(\dmini)=\hhi\}. 
    \end{aligned}
\end{equation}
In practice, the true functional form of $\hi(\cdot)$ is unknown to the researcher. However, we will show that estimation of treatment effects without direct knowledge of the exposure mapping’s functional form is feasible. 
\footnote{See \cite{verbitsky2012causal, delgado2015difference, clarke2017estimating, berg2020handling} for detailed discussions of causal inference settings with known exposure mappings.}\footnote{As will be discussed later in this section, to identify the treatment effects, we only require that a positive fraction of the population remains unaffected by spillovers.} 
To this end, let $\Si$ a binary indicator equal to $1$ if unit $i$ is ever subject to spillovers, and $0$ otherwise. We then impose the following assumption

\begin{assumption}[\textit{Irreversibility of Spillover}]\label{as_irreversibility_sp}
    $h_{it}(\mathbf{d}_{-i,t})=\mathbf{0} \text{ }\forall t<g$. For all $t \geq g$, $h_{it-1}(\mathbf{d}_{-i,t-1})\neq \mathbf{0}$ implies $\hi(\mathbf{d}_{-i,t})\neq \mathbf{0}$.
\end{assumption}

Analogously to \hyperref[as_irreversibility]{Assumption~\ref*{as_irreversibility}}, we require that no unit is subject to spillovers before the policy is implemented, and that once a unit is affected by spillovers, it remains affected in all subsequent periods.\footnote{One can interpret $\Si$ as the realization of an underlying latent variable--namely, the exposure mapping. In most settings, the exact functional form of $\hi(\cdot)$ is unknown to the researcher, but $\Si$ is observed. Drawing on the literature on random utility models, $\Si$ can be written as:

\begin{equation*}
    \begin{cases}
        \Si=1 & \text{if} \quad \hi(\mathbf{d}_{-i,t})\neq \mathbf{0}, \quad t\geq g,\\
        \Si=0 &  \text{if} \quad \hi (\mathbf{d}_{-i,t})=\mathbf{0}, \quad  t=1, \dots, T .
    \end{cases}
\end{equation*}
} 

We now focus on the missing data pattern. Let $A_{it}$ be a binary indicator equal to $1$ if unit $i$ is observed at time $t$, and $0$ otherwise. We define $A_{it-k,t} = A_{it-k} A_{it}$ to indicate whether unit $i$ is observed in both periods $t-k$ and $t$. Further, suppose there exists of a complete set of covariates, denoted by $\mathbf{X}_i$, with support denoted by $\chi = \mathrm{supp}(\mathbf{X}) \subseteq \mathbb{R}^K$ and rank equal to $\mathrm{rank}(\mathbf{X})=k$. The generalized propensity score can then be defined as $\P^s_{g}(\mathbf{X})=\P\left(G_{ig}=1\mid \mathbf{X}_i, G_{ig}+C_i=1, \Si=s, A_{it-k,t}=1\right)$ with $s\in\{0,1\}$, which represents the probability of being first treated in period $g$, conditional on being observed in both periods $t-k$ and $t$, on being subject or not to spillover, on observable characteristics, and either belonging to cohort $g$ or being never-treated.\footnote{When using not-yet-treated units as the comparison group, the generalized propensity score takes the form $\P^s_{g}(\mathbf{X}) = \P\left(G_{ig} = 1 \mid \mathbf{X}_i, G_{ig} + (1 - D_{is})(1 - G_{ig}) = 1, \Si=s, A_{it-k,t}=1 \right)$. Here, $\P_{g}(\mathbf{X})$ denotes the probability of being first treated in period $g$, conditional on observable characteristics and on either belonging to cohort $g$ (i.e., $G_{ig} = 1$) or being in the group not yet treated at time $s$.}
To define the parameters of interest, we impose two additional assumptions on the sampling process. Let $n_{t-k,t}^g$ denote the number of units treated in group $g$ and observed in both periods $t-k$ and $t$, and let $\Gamma^{g,ns}_{t-k,t}$ denote the subset of these units for which $\Si = 0$. We then assume:

\begin{assumption}[\textit{Sampling Process with Spillovers}]\label{as_RS}
 Conditional on $A_{it-k,t}=1$, for all $t=1,\dots, T$  and $k=1,\dots, t-1$: i) $Y_{it}$ is generated according to \hyperref[observed_y]{(\ref*{observed_y})}; ii) $\{ \X_i, D_{i1},\dots,D_{iT}\}^n_{i=1}$ is independent and identically distributed (iid); iii) For $g\in\G$ and $\hi(\dmini)=\hhi$, $\exists \text{ } \Gamma^{g,ns}_{t-k,t}\subseteq\{1,\dots, n_{t-k,t}^g\}$, such that $0<\mathrm{card}(\Gamma^{g,ns}_{t-k,t})/n_{t-k,t}^g<1$ and $Y_i(g, \hhi)\equiv Y_i(g, \0)$.
\end{assumption}

\begin{assumption}[\textit{Missing Trends at Random}]\label{as_Missing_trends}
For all $t=2,\dots, T$ and $k=1,\dots,t-1$,
\begin{equation}
    A_{it-k,t} \indep Y_{it}-Y_{it-k}, \X_i \mid \Gig, \Si.
\end{equation}
\end{assumption}

We observe that the third part of \hyperref[as_RS]{Assumption~\ref{as_RS}}, stating that in each group $g$ and time $t$ there exists a positive fraction of the treated population that is not affected by spillovers, is essential for identification of our treatment effects. In our empirical exercise, this implies that for every year and for each group we can find a subset of facilities that are not engaged in permit trading.\footnote{Note that, unlike Assumption 2 in \cite{Callaway2021}, we do not assume that the sample is drawn i.i.d. from the population.} \hyperref[as_Missing_trends]{Assumption~\ref*{as_Missing_trends}} posits that, conditional on treatment assignment and spillovers, the sampling mechanism is independent of the $k$-differenced outcomes, and the covariates. This assumption is reasonable in our empirical context, since missing data are most likely associated with firm closures, mergers, or acquisitions.

\subsection{Group-time average treatment effects under spillovers}

We now introduce two definitions of the average treatment effects on the treated. For $g \in \G$, the \textit{ATT without interference} at time $t$ for units first treated in period $g$ is

    \begin{equation}\label{ATT_0}
        ATT_0(g,t)= \E\left[Y_{it}\left(g,\mathbf{0}\right)-Y_{it}\left(0,\mathbf{0}\right)\mid G_{ig}=1, \Si = 0 \right].
    \end{equation} 
This parameter captures the average treatment effect  on the treated at time $t$ for units in group $g$ in the absence of spillovers. In other words, $ATT_0(g,t)$ isolates the direct effect of the treatment.
Another interesting parameter is given by the \textit{ATT with spillovers} in period $t$ for units first treated at time $g$
    \begin{equation}\label{ATT_S}
        ATT_S(g, t)= \E\left[Y_{it}(g,\hhi)-Y_{it}\left(0, \mathbf{0}\right)\mid G_{ig}=1, \Si =1\right].
    \end{equation}
In contrast to $ATT_0(g,t)$, $ATT_S(g,t)$ captures the average effect of the policy on the subpopulation first treated in period $g$ and subject to interference from other treated units. $ATT_S$ captures the impact of simultaneously receiving the treatment and transitioning from no exposure to being affected by spillovers. In other words, $ATT_S(g,t)$ quantifies the overall (average) effect of the policy, combining both the direct effect of the treatment and the indirect effect arising from spillovers generated by the intervention.\footnote{Note that the parameters in \hyperref[ATT_0]{(\ref*{ATT_0})} and \hyperref[ATT_S]{(\ref*{ATT_S})} are similar to the group-time average treatment effect in \cite{Callaway2021}.} Lastly, for $\hi(\dmini)=\hhi$, we define the \textit{average spillover effect on the treated} in period $t$ for units first treated at time $g$ and subject to spillover as

    \begin{equation}\label{A_SPILL}
        AST(g, t)= \E\left(Y_{it}\left(g, \hhi\right)-Y_{it}\left(g,\mathbf{0}\right)\mid \Gig=1, \Si=1\right).
    \end{equation}
The parameter $AST(g,t)$ captures the additional impact attributable to spillovers for units treated in group $g$.

\subsection{Identification}

To identify the parameters defined in \hyperref[ATT_0]{(\ref*{ATT_0})}–\hyperref[ATT_S]{(\ref*{ATT_S})}, we need to impose the following assumptions 

\begin{assumption}[\textit{Limited Treatment Anticipation}]\label{as_NA}
There is a known $\delta \geq 0$ such that 

\begin{equation*}
    \begin{aligned}
        &\E\left[\yit(g,\hhi)\mid \X_i, \Gig=1 \right]=\E\left[\yit(0,\0) \mid \X_i, \Gig=1 \right] \quad \text{and} \\
        & \E\left[\yit(g,\0 )\mid \X_i, \Gig=1\right]=\E\left[\yit(0,\0) \mid \X_i,  \Gig=1\right] \text{ a.s.} 
    \end{aligned}
\end{equation*}

for all $g\in \G,t \in \{1,\dots, T\}$  such that  $t<g-\delta$.
\end{assumption}

\begin{assumption}[\textit{Conditional Parallel Trends}]\label{as_PT} Let $\delta$ be as defined in \hyperref[as_NA]{Assumption~\ref*{as_NA}}. For each $g\in \G$ and $t \in \{2,\dots, T\}$ such that $ t\geq g-\delta$,
\begin{equation}
     \E\left( Y_{it} \left(0,\0 \right)- Y_{it-1}\left(0,\0 \right)\mid  \X_i,   G_{ig}=1\right)=\E\left(Y_{it}\left(0,\0 \right)- Y_{it-1}\left(0,\0 \right)\mid  \X_i,  C_i=0 \right) \quad \text{a.s.}
\end{equation}
\end{assumption}

\begin{assumption}[\textit{Overlap}]\label{as_Overlap} For each $t\in\{2,\dots, T\}$, $g \in \G $, $s\in\{0,1\}$ there exists some $\varepsilon>0$ s.t. $\P(\Gig=1)>\varepsilon$ and $\P^s_{g}(\mathbf{X})<1-\varepsilon$ a.s.
\end{assumption}

\hyperref[as_NA]{Assumption~\ref*{as_NA}} states that, on average, before treatment occurs, the treated and untreated potential outcomes of eventually treated (and eventually subject to spillover) units are the same in the pre-treatment period \cite[see also Assumption 3 in][]{Callaway2021}. \hyperref[as_PT]{Assumption~\ref*{as_PT}} generalises the classical Parallel Trends Assumption, stating that the evolution of untreated potential outcomes is mean independent of treatment assignment, once covariates are accounted for. Lastly, \hyperref[as_Overlap]{Assumption~\ref*{as_Overlap}} is the common overlap condition made in \cite{Callaway2021}. 
We now focus on identification of parameters in \hyperref[ATT_0]{(\ref*{ATT_0})}, since identification of \hyperref[ATT_S]{(\ref*{ATT_S})} can be obtained by a similar argument. Identification of the parameters in \hyperref[ATT_S]{(\ref*{ATT_S})} and \hyperref[A_SPILL]{(\ref*{A_SPILL})} is presented in Appendix~A. 
To this end, we follow the approach proposed by \cite{Bellegoetal2025} that consists of decomposing the long difference $\E\left[Y_{it} - Y_{ig-1} \mid G_{ig} = 1, \Si = 0 \right]$ into the sum of several $k$-period differences. To simplify the exposition, in what follows, we focus on identification (and estimation) based on 1-period differences decomposition. However, the results can be extended to the more general case of $k$-period differences, useful when the structure of unbalanced panels with missing data at random, and irregular gaps in the time series of each statistical unit. In the presence of irregular gaps, a Generalised Method of Moments based on a a combination of differences of various orders can be proposed. This case will be presented in Appendix~A. Consider:

\begin{equation}
\E\left[Y_{it} - Y_{ig-1} \mid G_{ig} = 1, \Si = 0 \right] = \sum_{\tau = g-\delta}^{t} \E\left[Y_{i\tau} - Y_{i\tau - 1} \mid \Gig=1 , \Si = 0\right].
\end{equation}

Accordingly, the $ATT_0(g, t)$ is obtained by summing the 1-period DiD components

\begin{equation*}
    \begin{aligned}
        ATT_0^{CD}(g,t)  = & \sum_{\tau=g-\delta}^{t} \Bigl(\E\left[Y_{i\tau}-Y_{i\tau-1}\mid \Gig=1, \Si=0 \right]-\E\left[Y_{i\tau}-Y_{i\tau-1}\mid C_i=1\right] \Bigl) \\
        = &  \sum_{\tau=g-\delta}^{t} \Bigl( \E\left[Y_{i\tau}-Y_{i\tau-1}\mid A_{i\tau-1,\tau}=1, G_{ig}=1, \Si=0\right] \\
        &- \E\left[Y_{i\tau}-Y_{i\tau-1}\mid A_{i\tau-1,\tau}=1, C_i=1\right] \Bigl).
    \end{aligned}
\end{equation*}
where the second equality holds under \hyperref[as_Missing_trends]{Assumption~\ref{as_Missing_trends}}.
Define the following weights

$$w_{i\tau-1,\tau}^G(g)= \frac{G_{ig}A_{i\tau-1,\tau}(1-\Si)}{\E_M\left[G_{ig}A_{i\tau-1,\tau}(1-\Si)\right]},$$

\noindent and

$$w_{i\tau-1,\tau}^C(g, \X)= \frac{\P_g^{0}(\X)C_i(1-\Si)A_{i\tau-1,\tau}}{1-\P_g^{0}(\X)}\Big/\E_M\left[\frac{\P_g^{0}(\X)C_i(1-\Si)A_{i\tau-1,\tau}}{1-\P_g^{0}(\X)}\right],$$
where $\E_M(\cdot)$ is the expectation with respect the mixture distribution $\F_M(\cdot)$ defined in Appendix A.

\begin{theorem}\label{Th1}
Under Assumptions~\hyperref[as_irreversibility]{\ref*{as_irreversibility}}--\hyperref[as_Overlap]{\ref*{as_Overlap}}, and for $g\in\G$ and $t\in\{2,\dots, T\}$, the long-term average treatment effect without interference in period $t$ is nonparametrically identified and given by
\begin{equation}
    ATT_0^{CD}(g,t)=\sum_{\tau=g-\delta}^t \Delta ATT_0(g,\tau),
\end{equation}

\sloppy \noindent
where $\Delta ATT_0(g,t)=\E_M\left[w_{i\tau-1,\tau}^G\left(g\right)\left(Y_{i\tau}-Y_{i\tau-1}\right)\right]-\E_M\left[w_{i\tau-1,\tau}^C\left(g, \X\right)\left(Y_{i\tau}-Y_{i\tau-1}\right)\right]$. 
\end{theorem}

\hyperref[Th1]{Theorem~\ref*{Th1}} states that the average treatment effect can be recovered by adding the 1-period difference in group-time average treatment effects without spillovers, measuring the increase of average treatment effect of group $g$ from period $t-1$ to $t$ in absence of spillovers. 
The proof is provided in Appendix~A. Similarly, identification of $ATT_S(g,t)$ and $AST(g)$ can be proved. 

Once the $ATT_0(g,t)^{CD}$ parameters are identified, these can be aggregated following the schemes proposed by \cite{Callaway2021} to explore heterogeneity along specific dimensions—such as how treatment effects vary with the length of exposure to the treatment. Alternatively, one can aggregate these causal parameters to summarize the overall treatment effect into a single parameter.

\subsection{Estimation and inference}

Results from \hyperref[Th1]{Theorem~\ref*{Th1}} suggest the following non-parametric estimator:

\begin{equation}\label{ChainedDID}
    \widehat{ATT}_0^{CD}(g,t)= \frac{1}{n_{t}^{g,c}}\sum_{i=1}^{n_{t}^{g,c}}\sum_{\tau=g-\delta}^t \{\hat w_{i\tau-1,\tau}^G(g) \left(Y_{i\tau}-Y_{i\tau-1}\right)-\hat w_{i\tau-1,\tau}^C(g, \X) \left(Y_{i\tau}-Y_{i\tau-1}\right)\},
\end{equation}

\noindent where $n_{t}^{g,c}$ denotes the sum of the number of treated units in cohort $g$ and the number of never treated units sampled in all periods between $g-1$ and $t$,

$$\hat w_{i\tau-1,\tau}^G(g)= \frac{G_{ig}A_{i\tau-1}A_{i\tau}(1-\Si)}{\frac{1}{n_{t}^{g,c}}\sum_{i=1}^{n_{t}^{g,c}}G_{ig}A_{i\tau-1}A_{i\tau}(1-\Si)},$$

\noindent and

$$\hat w_{i\tau-1,\tau}^C(g, \X)= \frac{\hat \P^0_g(\X)C_{i}(1-\Si)A_{i\tau-1}A_{i\tau}}{1-\hat \P^0_g(\X)}\Big/ \frac{1}{n_{t}^{g,c}}\sum_{i=1}^{n_{t}^{g,c}}\left[\frac{\hat \P^0_g(\X)C_i(1-\Si)A_{i\tau-1}A_{i\tau}}{1-\hat \P^0_g(\X)}\right],$$
where $\hat \P^s_g(\cdot)$ denotes an estimate of the propensity score (e.g., logit or probit), obtained in a first step.

Alternatively, if there is no anticipation (i.e., $\delta=0$) and covariates play no role, one can estimate the following regression using the full sample

\begin{equation}
    Y_{it}=\alpha_i+\eta_t+\sum_{g\in\G}\sum_{\tau=g}^T \beta_{g\tau} \left(G_{ig}\times (1-\Si)\times\i\{t=\tau\} \right)+ \sum_{g\in\G}\sum_{\tau=g}^T \gamma_{g\tau} \left(S_{i}\times G_{ig} \times\i\{t=\tau\}\right) +\varepsilon_{it},
\end{equation}
where $\alpha_i$ and $\eta_t$ denote unit and time fixed effects, respectively, and $\varepsilon_{it}$ is the error term. Under the same assumptions required for \hyperref[Th1]{Theorem~\ref*{Th1}} to hold, it can be shown that $\hat{\beta}_{g\tau}$ and $\hat{\gamma}_{g\tau}$  are unbiased, $\sqrt{n}$-consistent and asymptotically normally distributed estimators of the $ATT^{CD}_0(g,t)$ and  $ATT_S^{CD}(g,t)$, respectively \citep[we refer to][for details]{Bellegoetal2025}. Further, for computing standard errors of estimates, the multiplier bootstrap procedure proposed by \cite{Callaway2021} can still be adopted.

%----------------------------------------------------------------------------------------
%	MONTE CARLO SIMULATIONS
%----------------------------------------------------------------------------------------
\section{Monte Carlo experiments}\label{MonteCarlo}
\subsection{Design} 
We now examine the finite sample properties of our estimators for the group-time average treatment effects. Specifically, we compare the performance of our proposed DiD estimators for $ATT_0$, $ATT_S$ and $AST$ with that of the estimator proposed by \cite{Bellegoetal2025}, which ignores spillovers and instead targets $ATT$ (hereafter, the \textit{``conventional'' chained DiD}). We evaluate the estimators' performance in terms of average bias and root mean squared error (RMSE). All results are based on $2,000$ Monte Carlo simulations. We set $T = 10$, $\mathcal{G} =\{3, \dots, 8\}$, and allow the sample size to vary, $n = \{200, 500, 1000, 2000  \}$. Since the average treatment effects are estimated via the GMM approach, we report results obtained with both the identity matrix as weighting matrix (hereafter referred to as the \textit{id estimator}) and the two-step estimator of the optimal weighting matrix $\Omega$ (abbreviated as \textit{2-step estimator}). For ease of exposition, we report only results for $ATT(3,4)$, $ATT_0(3,4)$, $ATT_S(3,4)$ and $AST(3,4)$, using the ``not-yet-treated'' as the comparison group. We consider the following data-generating process (DGP) and define the untreated potential outcomes as
\begin{equation}
Y_{it}(0,\0) = \alpha_i +\eta_t + X_{it}\beta + u_{it}.    
\end{equation}
For units in group $g \neq 0$ and in their post-treatment periods (i.e., outcomes when $t \geq g$), we set
\begin{align}
Y_{it}(g,\hhi) &= \alpha_i + \eta_t +  X_{it}\beta + \delta_e + \gamma \Si +\nu_{it},
\end{align}
here, we set $\gamma = 1$ and define $\delta_e = e + 1 = t - g + 1$, where $e = t - g$ denotes the number of periods a unit has been under treatment by time $t$. $\Si$ is a binary indicator equal to $1$ if $h_{it}(\dmini) \neq \mathbf{0}$ and $0$ otherwise. Under this framework, if $h_{it}(\dmini) = \mathbf{0}$--i.e., in the absence of spillovers--then for any group and any post-treatment period, we have $ATT_0(g, t) = t - g + 1 = e + 1$. Finally, for units in group $g \neq 0$ that are subject to spillovers and observed in post-treatment periods, we obtain $ATT_S(g, t) = t - g + 2 = e + 2$.

In the above equations we set $\alpha_i \sim Unif(0.1,0.9)$, $\eta_t = t$, $\beta =  1$, $u_{it}\sim N(0,0.5)$, $\nu_{it}\sim N(0,0.5)$ and

\begin{equation*}
X_{it}=1+ 0.1\alpha_i + v_{it} , 
\end{equation*}
with $v_{it}\sim N(0,0.5)$. We further set the probability of belonging to group $g \in \G$ as

\begin{equation*}
\P(G_i = g|X_i) = \frac{e^{X_{i1}\kappa_g}}{\sum_{g=0}^G e^{X_{i1}\kappa_g}},    
\end{equation*}
where $\kappa_g = 0.5 \frac{g}{G}$ and $X_{i1}$ is the value of $X_i$ in period $t=1$. We then set the probability of being (ever) subject to spillover for treated units as

\begin{equation*}
\P(S_{i} = 1|\alpha_i, G_i=g) =\frac{e^{1 + 0.5\alpha_i}}{1 + e^{1 + 0.5\alpha_i} }.    
\end{equation*}
We finally assume missing data at random by setting

\begin{equation*}
    \P(A_{it}=1|\alpha_i) = \frac{e^{1 + 0.1\alpha_i t}}{1 + e^{1 + 0.1\alpha_i t} }.    
\end{equation*}
It can be proved that the above data-generating process satisfies Assumptions \hyperref[as_irreversibility]{\ref*{as_irreversibility}}-\hyperref[as_Overlap]{\ref*{as_Overlap}}.

\subsection{Monte Carlo results}
\hyperref[MCres]{Table~\ref*{MCres}} reports the results of the Monte Carlo simulations in a balanced panel setting, comparing the performance of the ``conventional'' chained DiD proposed by \cite{Bellegoetal2025}, which ignores spillover effects, with that of our proposed estimators for $ATT_0$, $ATT_S$ and $AST$ under varying sample sizes. All results are based on the inverse probability weighting estimand. As expected, the conventional chained DiD estimator exhibits a bias that does not diminish as the sample size increases. In contrast, the bias of our proposed estimators remains close to zero for all sample sizes, and their RMSE decreases as the sample grows. We also observe that the 2-step estimator has a larger RMSE than its ID counterpart, although this gap narrows as the sample size increases. \newline

\begin{table}[H]
\centering
\caption{Simulation results for $\widehat {ATT}^{CD}$ (ignoring spillovers), $\widehat {ATT}^{CD}_0$, $\widehat {ATT}^{CD}_S$ and $\widehat {AST}^{CD}$}
\label{MCres}
\begin{tabular}{clcccccccc}
  \hline
  \\[-1.8ex] 
 &  &  \multicolumn{2}{c}{$\widehat {ATT}^{CD}$} &  \multicolumn{2}{c}{$\widehat {ATT}^{CD}_0$} &  \multicolumn{2}{c}{$\widehat {ATT}^{CD}_S$}   &  \multicolumn{2}{c}{$\widehat {AST}^{CD}$} \\ 
  \cmidrule(lr){3-4}  \cmidrule(lr){5-6}  \cmidrule(lr){7-8}  \cmidrule(lr){9-10}
 \\[-1.8ex] 
N &  & $bias$ & $RMSE$ & $bias$ & $RMSE$ & $Bias$ & $RMSE$ & $Bias$ & $RMSE$\\
\\[-1.8ex] 
  \hline
200 & ID & 0.133 & 0.289 & 0.003 & 0.264 & 0.005 & 0.623 & 0.004 & 0.803 \\ 
  & 2-step &  0.132 & 0.342 & 0.007 & 0.377 & -0.219 & 1.196 & -0.039 & 2.574 \\ 
500 & ID & 0.122 & 0.202 & -0.000 & 0.167 & -0.022 & 0.381 & -0.016 & 0.638 \\ 
 & 2-step & 0.121 & 0.208 & -0.002 & 0.208 & -0.033 & 0.814 & -0.003 & 0.717 \\ 
1000 & ID & 0.122 & 0.167 & -0.003 & 0.120 & 0.001 & 0.249 & 0.001 & 0.456 \\ 
 & 2-step & 0.123 & 0.169 & -0.004 & 0.137 & 0.010 & 0.459 & 0.002 & 0.471 \\ 
2000 & ID & 0.126 & 0.149 & 0.002 & 0.081 & 0.001 & 0.179 & -0.007 & 0.303 \\ 
 & 2-step & 0.126 & 0.149 & -0.000 & 0.086 & -0.006 & 0.239 & -0.008 & 0.305 \\ 
\\[-1.8ex] 
\hline
\end{tabular}
\floatfoot{\footnotesize{\textbf{Notes}: This table reports the Monte Carlo results for the estimator of the $ATT^{CD}(g,t)$ by \cite{Bellegoetal2025} and for our proposed estimators of $ATT^{CD}_0(g,t)$, $ATT^{CD}_S(g,t)$ and $AST^{CD}(g,t)$ in the setup with $T = 10$ and parameter of interest $ATT^{CD}(3,4)$. Each Monte Carlo simulation uses $2,000$ bootstrap replications. In these experiments we rescale $\P(S_i=1|\alpha_i, G_i=g)$ so that the average probability is $0.50$. Rows labelled ``ID'' present results obtained via the GMM approach with the identity weighting matrix, while rows labelled ``2-step'' correspond to the GMM approach with the two-step estimator for the optimal weighting matrix.}}
\end{table}

In \hyperref[MCres2]{Table~\ref*{MCres2}}, we consider 
an unbalanced panel (Panel~(a)), and a severely unbalanced panel where, on average, only half of the sample is observed (Panel~(b)). Specifically, we rescale the selection probabilities so that $\bar{\P}(A_{it}=1|\alpha_i)=0.70$ and $\bar{\P}(A_{it}=1|\alpha_i)=0.50$, respectively, while we rescale the share of units affected by spillovers so that $\bar{\P}(S_{i}=1|\alpha_i)=0.50$. In an additional experiment, we take a balanced panel and vary the share of units affected by spillovers by rescaling $\P(S_i=1|\alpha_i, G_i=g)$ so that the average probability is $0.70$ (Panel~(c)), or $0.30$ (Panel~(d).  These cases are particularly relevant, as in our empirical application approximately 43 percent of the complete panel is missing, and around 40 per cent of the treated units are subject to spillovers.
In the case of severe unbalancedness--where only half of the sample is observed on average--the RMSE of both $ATT_0$, $ATT_S$ increase substantially, particularly that of the 2-step estimators. The $AST$ displays a large increase in bias and RMSE, probably linked to the small size of the sample used for estimation.

When a large share of units is affected by spillovers (Panel~(c)), the performance of the conventional chained DiD estimator is poor, while as expected the bias and RMSE of our estimators (both ID and 2-step) are very small. When, on average, thirty per cent of the treated population is affected by spillovers (Panel~(d)), the bias and RMSEs of the conventional chained DiD slightly improves, while the RMSEs of both the $\widehat {ATT}^{CD}_S$ and $\widehat {AST}^{CD}$ estimators increases. 
This arises because the number of units affected by spillovers is relatively small so these parameter are imprecisely estimated. In this setting, the conventional DiD performs somewhat better than before, since neglecting spillovers matters less when only about $30\%$ of treated units are affected.  

Overall, the simulation results indicate that failing to account for spillovers leads to severe bias and unreliable inference, that our proposed estimators perform well in a balanced and unbalanced setting, with the ID approach preferable under greater unbalancedness. Similar results are found for the other group-time average treatment effect estimators. These results are available upon request.

\begin{table}[H]
\centering
\caption{Simulation results for $ATT^{CD}$ (ignoring spillovers), $ATT^{CD}_0$, $ATT^{CD}_S$ and $AST^{CD}$ with varying degrees of unbalancedness and fraction of units subject to spillovers}
\label{MCres2}
\begin{tabular}{lcccccccc}
  \hline
  \\[-1.8ex] 
 &  \multicolumn{2}{c}{$ATT^{CD}$} &  \multicolumn{2}{c}{$ATT^{CD}_0$} &  \multicolumn{2}{c}{$ATT^{CD}_S$}   &  \multicolumn{2}{c}{$AST^{CD}$} \\ 
 \cmidrule(lr){2-3}  \cmidrule(lr){4-5}  \cmidrule(lr){6-7}  \cmidrule(lr){8-9} 
\\[-1.8ex] 
 &  $bias$ & $RMSE$ & $bias$ & $RMSE$ & $Bias$ & $RMSE$ & $Bias$ & $RMSE$\\ 
 \hline \\[-1.8ex] 
 &  \multicolumn{8}{c}{(a): $\bar \P(A_{it} = 1|\alpha_i)=0.70$}\\
  \hline
ID &  0.130 & 0.035 & 0.004 & 0.018 & -0.004 & 0.098 & 0.003 & 0.352  \\ 
2-step & 0.121 & 0.039 & 0.000 & 0.026 & -0.003 & 0.402 & -0.009 & 0.795  \\ 
\\ 
\hline \\[-1.8ex] 
 &  \multicolumn{8}{c}{(b): $\bar \P(A_{it} = 1|\alpha_i)=0.50$}\\
  \hline
ID & 0.125 & 0.047 & -0.001 & 0.035 & -0.026 & 0.237 & -0.280 & 0.483  \\ 
2-step & 0.124 & 0.065 & 0.000 & 0.062 & 0.008 & 0.804 & -0.283 & 1.534 \\ 
\\ 
\hline \\[-1.8ex] 
 &  \multicolumn{8}{c}{(c): $\bar \P(S_{i} = 1|\alpha_i)=0.70$}\\
 \hline
ID &  0.176 & 0.044 & 0.001 & 0.015 & -0.000 & 0.048 & 0.001 & 0.138 \\ 
2-step & 0.176 & 0.045 & -0.001 & 0.019 & 0.004 & 0.106 & 0.003 & 0.147  \\
\\ 
\hline \\[-1.8ex] 
 &  \multicolumn{8}{c}{(d): $\bar \P(S_{i} = 1|\alpha_i)=0.30$}\\
  \hline
ID &  0.078 & 0.018 & 0.002 & 0.013 & 0.002 & 0.124 & 0.009 & 0.374  \\ 
2-step & 0.077 & 0.019 & 0.002 & 0.020 & 0.023 & 0.578 & 0.005 & 0.443 \\ 
\\[-1.8ex] 
\hline 
\end{tabular}
\floatfoot{\footnotesize{\textbf{Notes}: This table reports the Monte Carlo results for the estimator of the $ATT^{CD}(g,t)$ by \cite{Bellegoetal2025} and for our proposed estimators of $ATT^{CD}_0(g,t)$, $ATT^{CD}_S(g,t)$ and $AST^{CD}(g,t)$ in the setup with $n=1,000$, $T = 10$ and parameter of interest $ATT^{CD}(3,4)$. Each Monte Carlo simulation uses $2,000$ bootstrap replications. Rows labelled ``ID'' present results obtained via the GMM approach with the identity weighting matrix, while rows labelled ``2-step'' correspond to the GMM approach with the two-step estimator for the optimal weighting matrix.}}
\end{table}

\section{Environmental policy and firm performance in Europe}\label{empirical}

Introduced in 2005, the EU ETS is the world's largest carbon market of the EU's climate strategy. According to the Emissions Trading Directive, participation in the EU ETS is mandatory for all plants with a rated thermal input exceeding 20MWh. In addition, the scheme covers large installations engaging in activities that are intensive in CO\textsubscript{2} emissions such as mineral oil refineries, coke ovens, iron and steel, and factories producing cement, glass, lime, bricks, ceramics, and pulp and paper. Recent years have seen an expansion of sectoral coverage to include airlines and aluminium manufacturing installations. Allowances, known as European Union Allowances (EUAs), are allocated to installations, ensuring that the total amount of emissions stays within the cap. Each EUA permits the emission of one tonne of CO\textsubscript{2}, and they are distributed at the installation level (i.e., individual plants). At the end of each year, regulated plants are required to \textit{surrender} one EUA for each tonne of CO\textsubscript{2} equivalent they have emitted over the year. Emitters can decide how much to pollute: if their emissions exceed the EUAs they hold, they must purchase additional EUAs from those with a surplus. This creates an incentive for companies to reduce emissions, for instance, through technological innovations, allowing them to sell excess EUAs for profit. Over time, the annual cap has been gradually lowered to ensure a consistent decline in overall emissions. Since its start, the EU ETS has been subject to several revisions, in order to boost its effectiveness. During Phase II (2008-2012), emission caps were tightened, and the system was expanded to include more sectors and greenhouse gases. Phase III (2013-2020) introduced a single, EU-wide cap on emissions, increased the auctioning of allowances, and implemented protections against carbon leakage. 
These adjustments have been made with the intent to make the EU ETS a central tool in the fight against climate change, aligning with the EU’s climate goal of achieving net-zero emissions by 2050.  

\subsection{Background literature on the impact of EU ETS}
A growing body of literature has attempted to estimate the causal impact of the above scheme on a range of outcomes for regulated firms, including carbon emissions, economic performance, and technological innovation. Most of these studies compare outcomes of regulated and non-regulated firms, before and after the implementation of the EU ETS, often focusing on specific countries. Using a Difference-in-Differences matching estimator applied to a panel data of German firms, \cite{Petrick2014} find that the EU ETS led to an approximate 25\% reduction in emissions at regulated firms. This reduction was achieved primarily through improvements in energy efficiency and reductions in the use of natural gas and petroleum products. Applying a similar approach,  \cite{jurate2016}  examine the impact of the EU ETS on CO\textsubscript{2} emissions and economic performance in Lithuania firms between 2005 and 2010. The study finds no significant reductions in emissions and only modest improvements in emission intensity during 2006-2007. \cite{calel2016} explores the impact of the cap and trade policy on technological change, showing that the EU ETS has resulted in a 10\% increase in low-carbon innovation amongst regulated plants. Similarly, \cite{calel2020} investigates the effects of the European carbon market on firm-level carbon intensity, patenting and R\&D expenditure in the UK from 2005 to 2012. Using a matched data set and applying the Hodges–Lehmann estimator to account for censoring in patenting and R\&D data, the study finds that although the EU ETS stimulated low-carbon innovation, it did not lead to reductions in carbon intensity. \cite{Dechezlepretre2023} analyse the impact of the European carbon market on facility-level emissions and firm-level economic performance across four European countries. Their Difference-in-Differences analysis shows a 10\% decline in carbon emissions in the years following the policy implementation, with no significant adverse effects on profits or employment. Finally, \cite{Colmeretal2024} use facility-level data from approximately 9,500 French manufacturing firms to examine the policy effectiveness. Their results show that ETS-regulated facilities reduced emissions by an average of 8-12\% relative to a matched group of unregulated facilities. The authors also explore the underlying mechanisms behind these findings, showing that the main primary driver of the observed emissions reductions is treated firms' efforts to lower the carbon intensity of production by upgrading their capital stock.

\subsection{Data}

Our first source of data is the E-PRTR database, established by the European Commission in July 2000 to collect comparable emission data from large industrial facilities across Europe.\footnote{\url{https://www.eea.europa.eu/data-and-maps/data/member-states-reporting-art-7-under-the-european-pollutant-release-and-transfer-register-e-prtr-regulation-23/european-pollutant-release-and-transfer-register-e-prtr-data-base.}} To be included in the database, a facility must operate in specific industrial sectors and exceed certain production and emissions thresholds.\footnote{See Annex I and II of the E-PRTR Regulation.} The database includes information on greenhouse gas emissions from $8,945$ unique facilities across 32 countries, for the years 2001, 2004, and 2007 to 2017.  We aggregate emission data for 6 main greenhouse gases\footnote{These are CO\textsubscript{2}, CH\textsubscript{4}, HFCS, N\textsubscript{2}O, PFCS and SF\textsubscript{6}.} into an overall measure of emissions that exploits information on the global warming potential of each pollutant and is expressed in 1,000 of tons of CO\textsubscript{2} equivalent, and drop observations with missing values. We refer the reader to Appendix B for a detailed description of the construction of our emissions measure. In addition to emissions data, the E-PRTR provides company-level information, including the country, the city and the sector where it operates. We then collected information on facilities regulated by the EU ETS drawing from the CITL database.\footnote{\url{https://www.euets.info}.} This database covers the period from 2005 to 2023 and includes information on $15,087$ installations belonging to $8,373$ firms. From the CITL emissions data, we extracted information on account opening dates, free emissions allowances, verified emissions, and transactions (i.e., the number of permits acquired and transferred). Accounts are central to participation in the EU ETS, as only installations with an active account can hold and trade allowances \citep{abrell2024transactions}. The account holder in the CITL is the operator (a firm or a facility) that owns or manages that installation. The account opening date indicates when an installation first became subject to the EU ETS. Free emissions allowances refer to CO\textsubscript{2} permits allocated to firms at no cost.  
Finally, transaction data provide information on the number of permits traded, which we use to infer whether an installation is subject to spillovers.\newline

To assess whether the introduction of the EU ETS led to a reduction in CO\textsubscript{2} emissions in Europe, we combine the CITL and E-PRTR data sets at the \textit{facility} level, and use data from CITL to identify those that have one or more installations treated under the EU ETS. We observe that not all CITL installations appear in the E-PRTR database because E-PRTR reporting depends on pollutant-specific release thresholds, whereas the EU ETS covers installations from specific sectors or exceeding certain capacity thresholds, even if their emissions remain below E-PRTR reporting limits. A key challenge in merging the E-PRTR and CITL data sets is that, while the CITL database contains information on the E-PRTR facility unique identifier associated to each installation, this is non-missing only for around 30 per cent of observations in the database. 
To join the two data sets, we implement a three-step matching procedure. In the first step, we carry an exact linkage between the two data sets using the E-PRTR unique identifier. In a second step, we apply the probabilistic record linkage approach proposed by \cite{ENAMORADO_FIFIELD_IMAI_2019} to match the facility name in the E-PRTR data set with the account name in the CITL data set, using as auxiliary information the operating country of the facility. The result from these two steps yields to identify  2,693 facilities from the CITL data set that are also present in the E-PRTR database. 
For the remaining set of unmatched facilities we exploit the fact that the CITL database contains the Orbis firm unique identifier (\texttt{BvDID number}) we take the name, city and operating country of all facilities in the E-PRTR data set and perform a search using Orbis’s batch function to retrieve their \texttt{BvDID number}.\footnote{Orbis is a comprehensive global database on private companies. For more information, see \href{https://orbis.bvdinfo.com/}{\textbf{https://orbis.bvdinfo.com/}}.} Accordingly, we use the BvDID number to link records from the CITL and E-PRTR data sets at the company level and identify treated firms. By doing this, we recover another 1,087 facilities that are present in both data sets.\newline

Once successfully joined the two data sets,\footnote{The final dataset is publicily available at \url{https://github.com/tosetel/EPRTR_EUETS_joined}} we define treatment status of a facility based on the presence of one or more of its installations within the CITL data set. To determine the entry year into the EU ETS, we use information on account openings, and assume that a facility joined the scheme in the year it opened its first EU ETS account. In contrast, the exit of the firm is identified by observing whether the firm continues to appear in the E-PRTR but has no longer any active accounts. We define a regulated (or treated) facility as subject to spillovers in a specific year if in that year it either acquires or transfers at least one emission permit to any other installation in the data set.
 Finally, we trim the top and bottom $5\%$ of the CO\textsubscript{2} emissions distribution to reduce the influence of extreme outliers. After this cleaning and selection procedure, our final data set consists of $43,570$ facility-year observations from $7,346$ unique facilities. Of these, $4,710$ (approximately $64\%$) were never treated and $2,636$ ($36\%$) received the treatment at some point. Finally, around 41 per cent of regulated facilities are subject to spillovers at some point in the treatment period, implying that they trade with other plants in the data set by selling or acquiring permits.\newline

\hyperref[panel_balance]{Table~\ref*{panel_balance}} provides descriptive statistics on panel balance, displaying the number of units observed over varying lengths of time and distinguishing between never treated and treated groups. While the distribution includes a significant share of short panels, particularly the 2,724 units observed for at most two periods, data also include a substantial number of long panels, with $1,857$ companies observed for at least 11 years.
\hyperref[staggered_tr_adop]{Table~\ref*{staggered_tr_adop}} in Appendix C complements this table by detailing the staggered implementation of the policy. The data indicate that the EU ETS was first implemented in 2005, with new firms entering the scheme every year up to 2020. Adoption was strongly concentrated in the early years: in 2005 alone, $1,625$ firms entered the scheme, accounting for $22.12\%$ of the entire sample. When combined with the 427 firms treated in 2006, nearly $30\%$ of all firms received treatment within the first two years. In later years, however, treatment adoption slowed considerably, with only a small number of new firms entering annually after 2014. Given the sparse number of newly treated firms in the years following 2006, we redefine the treatment start year as the first year of the EU ETS phase in which a firm entered the scheme, rather than the exact calendar year of entry. Such aggregation, addresses the purely statistical concern arising from the small number of late adopters, and accounts for possible anticipatory behaviour among firms that entered later within the same phase. For example, a firm opening an account between 2005 and 2007 is considered to have been first treated in 2005, the start of Phase I of the EU ETS. This leads to around 82 per cent of treated units first treated in 2005, with the remaining treated units equally split between 2008 and 2013.
Although we treat firms as being at an absorbing state once they enter treatment for our analysis, it is worth noting that some firms, in fact, exit treatment. \hyperref[leavers]{Table~\ref*{leavers}} shows that, between 2007 and 2017, 202 facilities left treatment status. Nevertheless, following \cite{Sun2021}, we assume that once a firm undergoes the policy, its behaviour is permanently affected. 
\begin{table}[H]
\begin{center}
\caption{\title{Descriptive Statistics on Panel Balance} \label{panel_balance}}
\begin{tabular}{c*{4}{c}}
\toprule
   \textbf{N. Periods}  &\textbf{Never Treated}&\textbf{Treated}&\textbf{Total}&\textbf{Percent}\\
\cline{1-5}
\\[-1.8ex] 
1           &        1120&         444&        1564&       21.29\\
2           &         843&         317&        1160&       15.79\\
3           &         253&         133&         386&        5.25\\
4           &         260&         132&         392&        5.34\\
5           &         239&         113&         352&        4.79\\
6           &         183&         103&         286&        3.89\\
7           &         202&         111&         313&        4.26\\
8           &         178&         111&         289&        3.93\\
9           &         200&         133&         333&        4.53\\
10          &         246&         168&         414&        5.64\\
11          &         497&         427&         924&       12.58\\
12          &         233&         216&         449&        6.11\\
13          &         256&         228&         484&        6.59\\
\cline{1-5}
Total       &        4710&        2636&        7346&      100.00\\
\bottomrule
\end{tabular}
\end{center}
\end{table}

\begin{figure}[H]
   \centering
    \caption{Average Trajectory CO\textsubscript{2} Equivalent [1,000 tons/year]} \label{average_trajectory}
    \begin{subfigure}{0.49\linewidth}
        \centering
        \includegraphics[width=\textwidth]{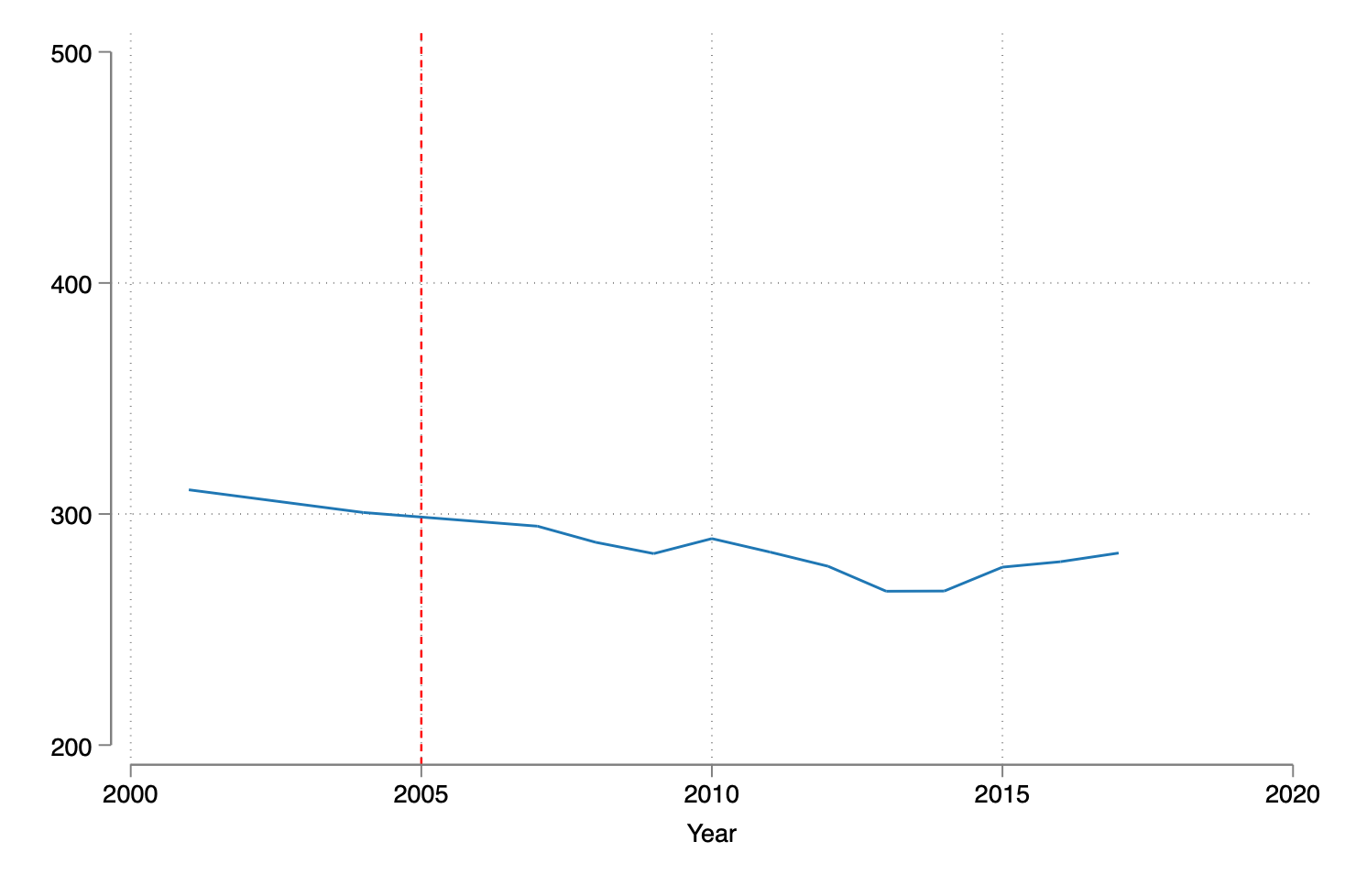}
       \caption{Overall}
        %\label{fig:first}
  \end{subfigure}
  \hfill
  \begin{subfigure}{0.49\linewidth}
     \centering
     \includegraphics[width=\textwidth]{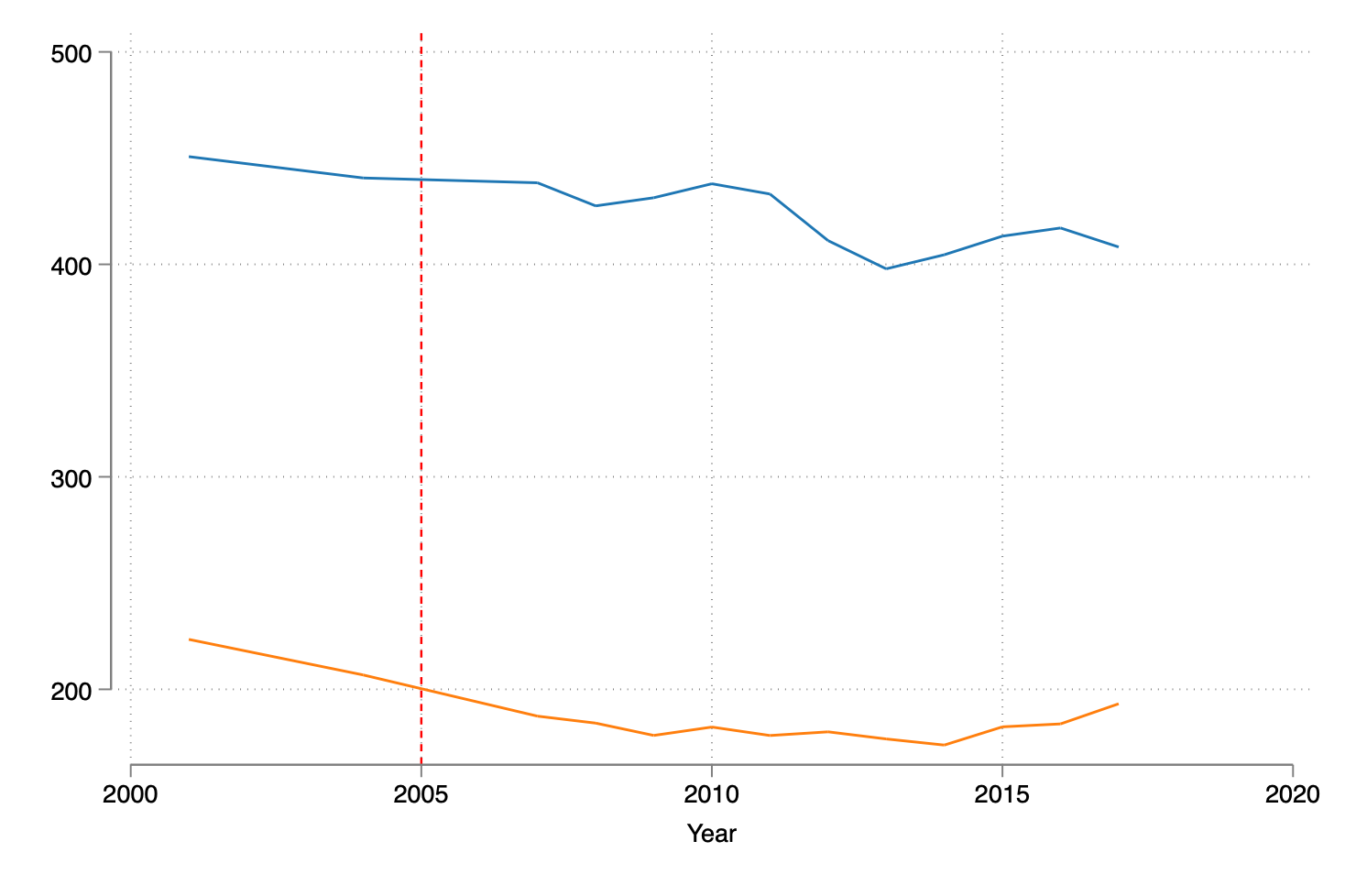}        \caption{Treated vs never treated}
        %\label{fig:second}
    \end{subfigure}
    \label{fig:TS_plot}
      \floatfoot{\footnotesize{\textbf{Notes}: This figure shows the temporal evolution of average CO\textsubscript{2} emissions, both overall (Panel (a)) and separately for treated and never-treated firms (Panel (b)). In Panel (b), the average trajectory for treated firms is shown in blue, while the trajectory for never-treated firms is shown in orange.}}
\end{figure}

\hyperref[average_trajectory]{Figure~\ref*{average_trajectory}} shows the temporal evolution of average CO\textsubscript{2}, both overall (Panel~(a)) and for treated versus never treated facilities (Panel~(b)). The left panel shows that prior to 2005, emissions appear relatively stable with minor fluctuations. After the introduction of the EU ETS, there is a more sustained decline in emissions, particularly from around 2007 onward. The right panel shows that, as expected, regulated facilities have a much higher average pollution than non-regulated ones. However, this gap narrows following the implementation of the scheme, particularly after 2007. Overall, this figure suggests that the EU ETS may indeed have led to a reduction in overall emissions. 

\hyperref[diff_means3]{Table~\ref*{diff_means3}} shows descriptive statistics on annual CO\textsubscript{2}-equivalent emissions (in 1,000 tons per year) before and after the EU ETS implementation, separately for never treated firms and for treated firms--both in the full sample and split by exposure to potential spillovers. Overall, treated facilities consistently emit more than never treated units, although both groups experience a statistically significant decline in emissions after 2005. On the right panel of the table we observe a substantial heterogeneity among the regulated facilities: emissions reduce by $32.98$ (significant at $1\%$) for treated facilities not subject to spillovers, while slightly increase (with a statistically  insignificant effect) for those subject to spillovers. This indicates that following the policy's introduction, treated plants exhibit a pronounced decline in emissions only among those that did not trade, supporting the view that while the EU ETS contributes to emission reductions, permit trading may attenuate the observable treatment effects. 

\begin{table}[htbp]
\begin{center}
\caption{\title{Descriptive Statistics on CO\textsubscript{2} Equivalent [1,000 tons/year] -- Pre vs Post EUETS} \label{diff_means3}}
\resizebox{\linewidth}{!}{\begin{tabular}{l*{14}{c}}
\toprule
 \vspace{0.00001cm}\\
& \multicolumn{4}{c}{\textbf{Never Treated}} & \multicolumn{10}{c}{\textbf{Treated}} \\
\cmidrule(lr){2-5} \cmidrule(lr){6-15}
& \textbf{Pre EUETS} & \textbf{Post EUETS} & \textbf{Diff.} & \textbf{Pval} & \textbf{Pre EUETS} & \multicolumn{9}{c}{\textbf{Post EUETS}} \\
\cmidrule(lr){2-5} \cmidrule(lr){6-6} \cmidrule(lr){7-15}
            &            &            &            &            &            &            &            &            &            &            &            &            &            &            \\
            &&&&&       &\textbf{Full Sample}&\textbf{Diff.}&\textbf{Pval}&\textbf{No Spillover}&\textbf{Diff.}&\textbf{Pval}&\textbf{Spillover}&\textbf{Diff.}&\textbf{Pval}\\
\cmidrule(lr){7-9} \cmidrule(lr){10-12} \cmidrule(lr){13-15}
\vspace{0.00001cm}\\
Mean        &      214.33&      181.72&      -32.61&        0.00&      445.02&      420.39&      -24.63&        0.03&      412.04&      -32.98&        0.00&      446.45&        1.43&        0.91\\
SD/SE       &      365.63&      349.83&        7.06&           .&      469.74&      443.40&       11.20&           .&      439.33&       11.36&           .&      454.95&       12.90&           .\\
N Obs.           &     3,011&    22,576&           .&           .&     1,951&    16,032&           .&           .&    12,144&           .&           .&     3,888&           .&           .\\
\bottomrule
\end{tabular}}
\end{center}
\end{table}

\section{Empirical Results}\label{results}    

\subsection{Main results}

To estimate the impact of the EU ETS, we apply the method presented in \hyperref[methods]{Section~\ref{methods}} using the not-yet-treated as the comparison group. To report results we adopt the aggregation schemes proposed by \cite{Callaway2021}. In particular, let $e=t-g$, we focus on aggregation schemes of the form  $\theta_{es}(e) =\sum_{g\in\G}\mathbbm{1}\{g+e\leq T\} \P\left(G=g|G+e\leq T\right)ATT^{CD}(g,g+e)$ which address the question: ``How does the average treatment effect vary with the duration of exposure to the treatment?''. We apply the same aggregation to $ATT_0^{CD}$ and $ATT_S^{CD}$.\newline

\begin{figure}[H]
    \centering
    \caption{Chained DiD Estimates}

    % Row 1 title
    \vspace{0.2cm}
    {\textbf{\large $\widehat{ATT}^{CD}$} \par}

    \begin{subfigure}{0.49\linewidth}
        \centering
        \includegraphics[width=\textwidth]{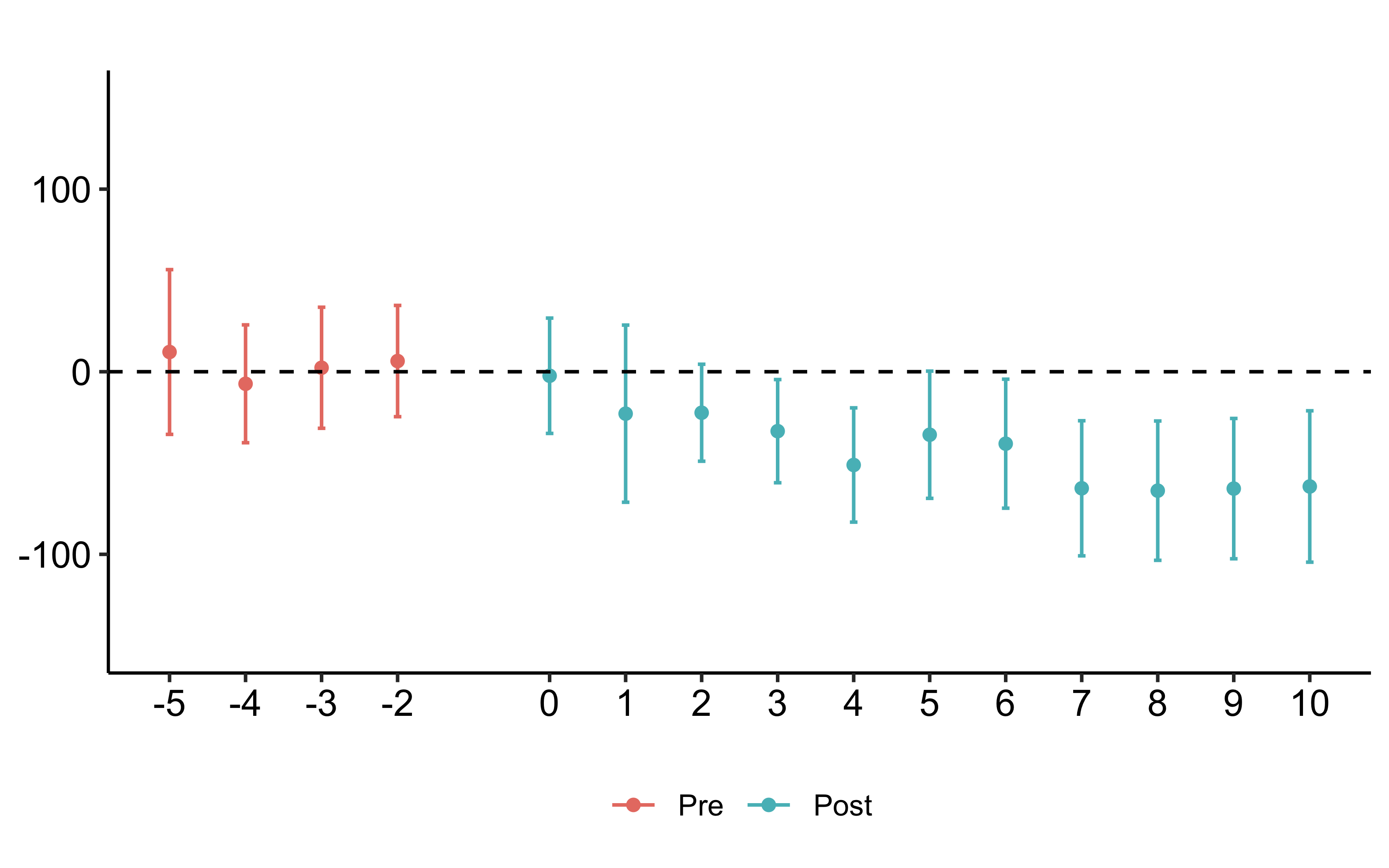}
        \caption*{(a) Without Covariates}
    \end{subfigure}
    \hfill
    \begin{subfigure}{0.49\linewidth}
        \centering
        \includegraphics[width=\textwidth]{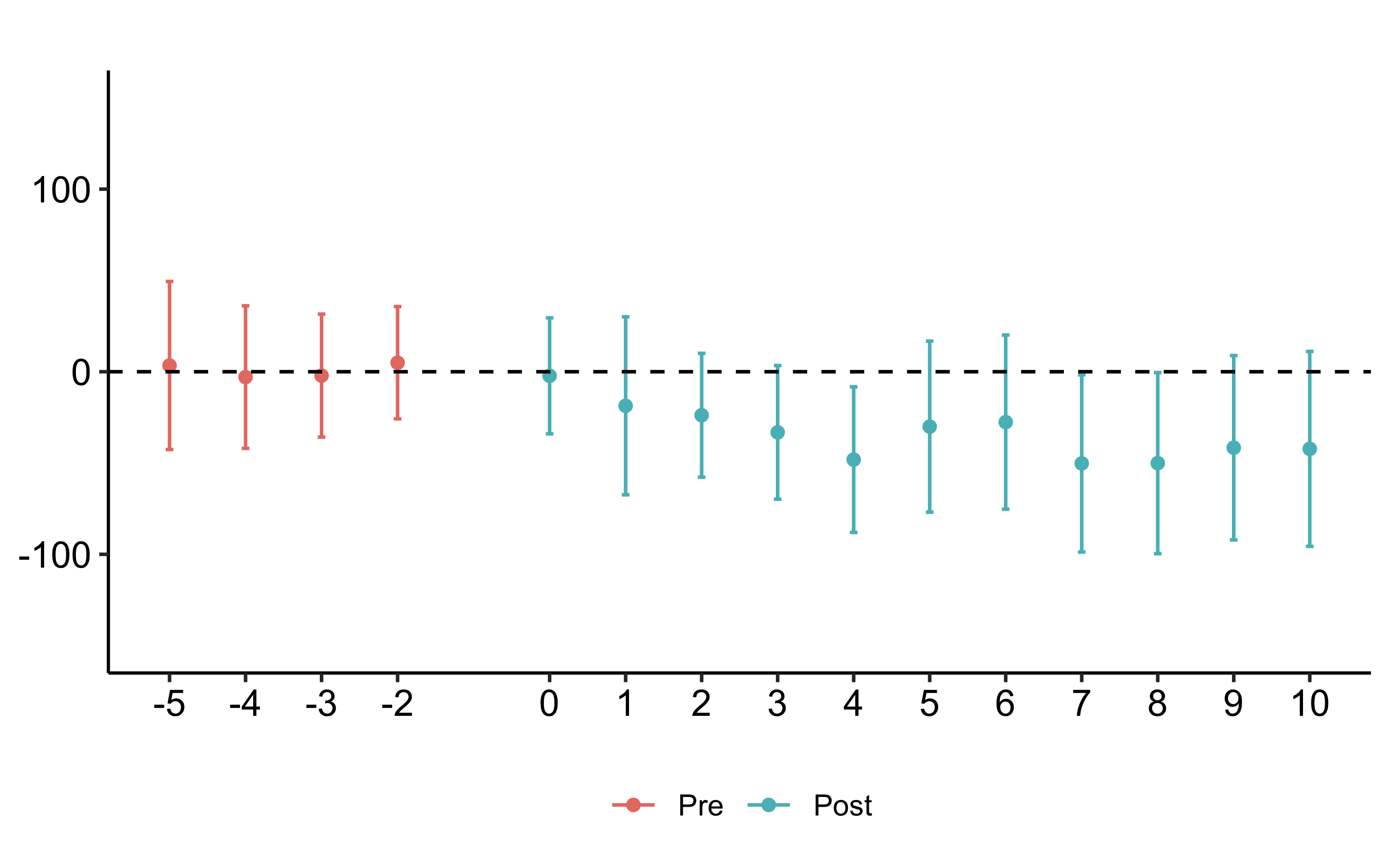}
        \caption*{(b) With Covariates}
    \end{subfigure}

    \vspace{0.6cm}

    % Row 2 title
    {\textbf{\large $\widehat{ATT}^{CD}_0$} \par}

    \begin{subfigure}{0.49\linewidth}
        \centering
        \includegraphics[width=\textwidth]{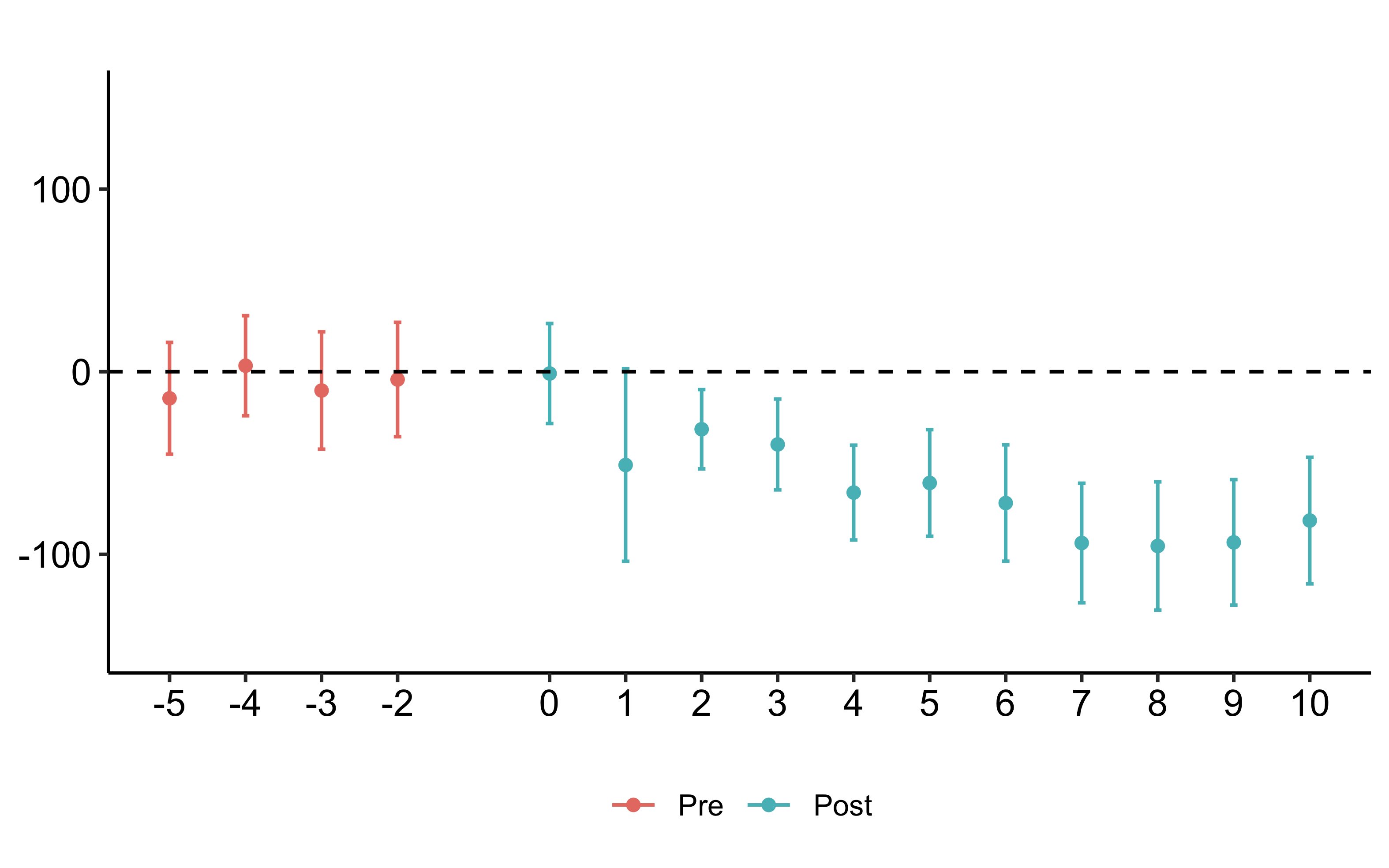}
        \caption*{(a) Without Covariates}
    \end{subfigure}
    \hfill
    \begin{subfigure}{0.49\linewidth}
        \centering
        \includegraphics[width=\textwidth]{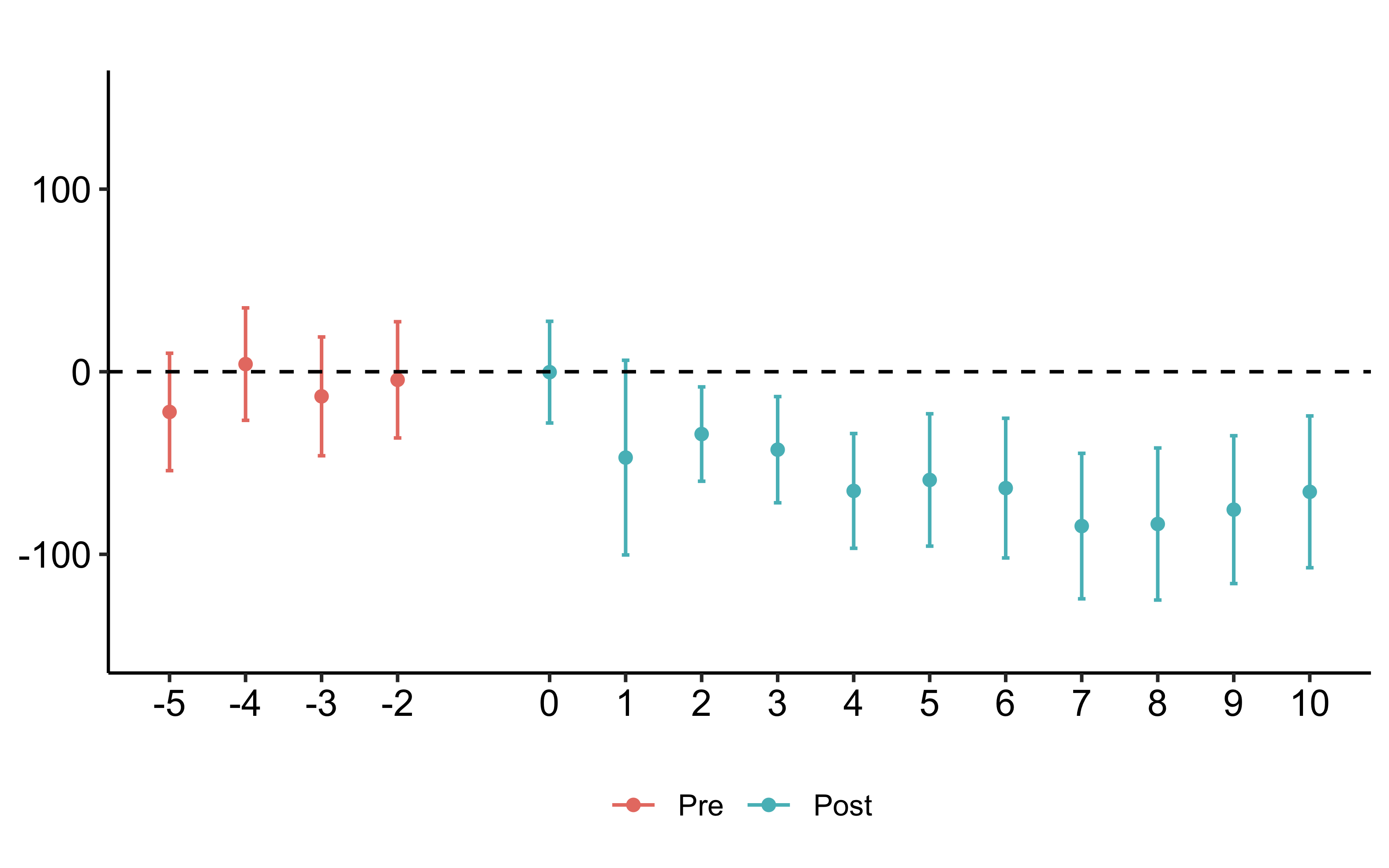}
        \caption*{(b) With Covariates}
    \end{subfigure}

    \vspace{0.6cm}

    % Row 3 title
    {\textbf{\large $\widehat{ATT}^{CD}_S$} \par}

    \begin{subfigure}{0.49\linewidth}
        \centering
        \includegraphics[width=\textwidth]{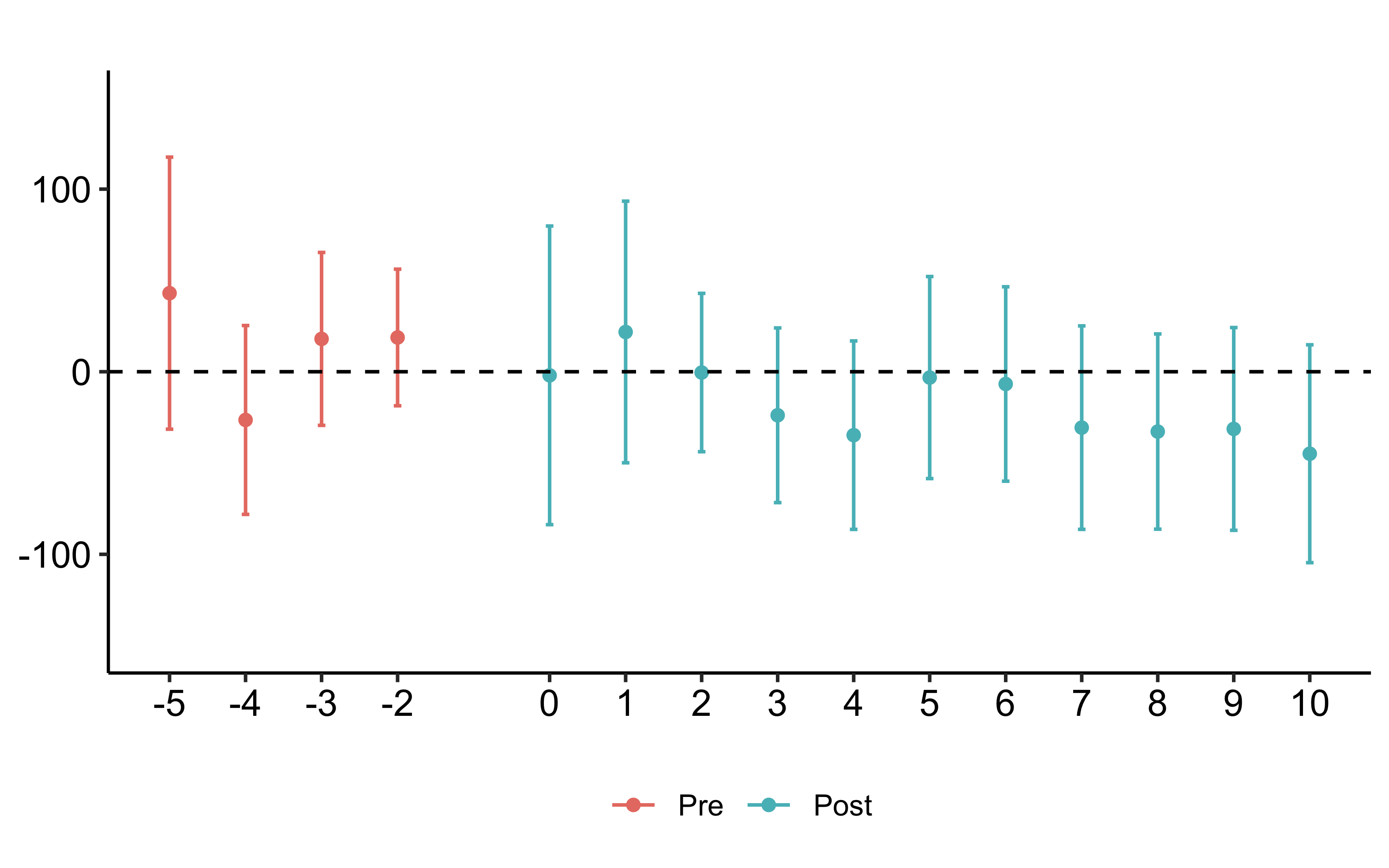}
        \caption*{(a) Without Covariates}
    \end{subfigure}
    \hfill
    \begin{subfigure}{0.49\linewidth}
        \centering
        \includegraphics[width=\textwidth]{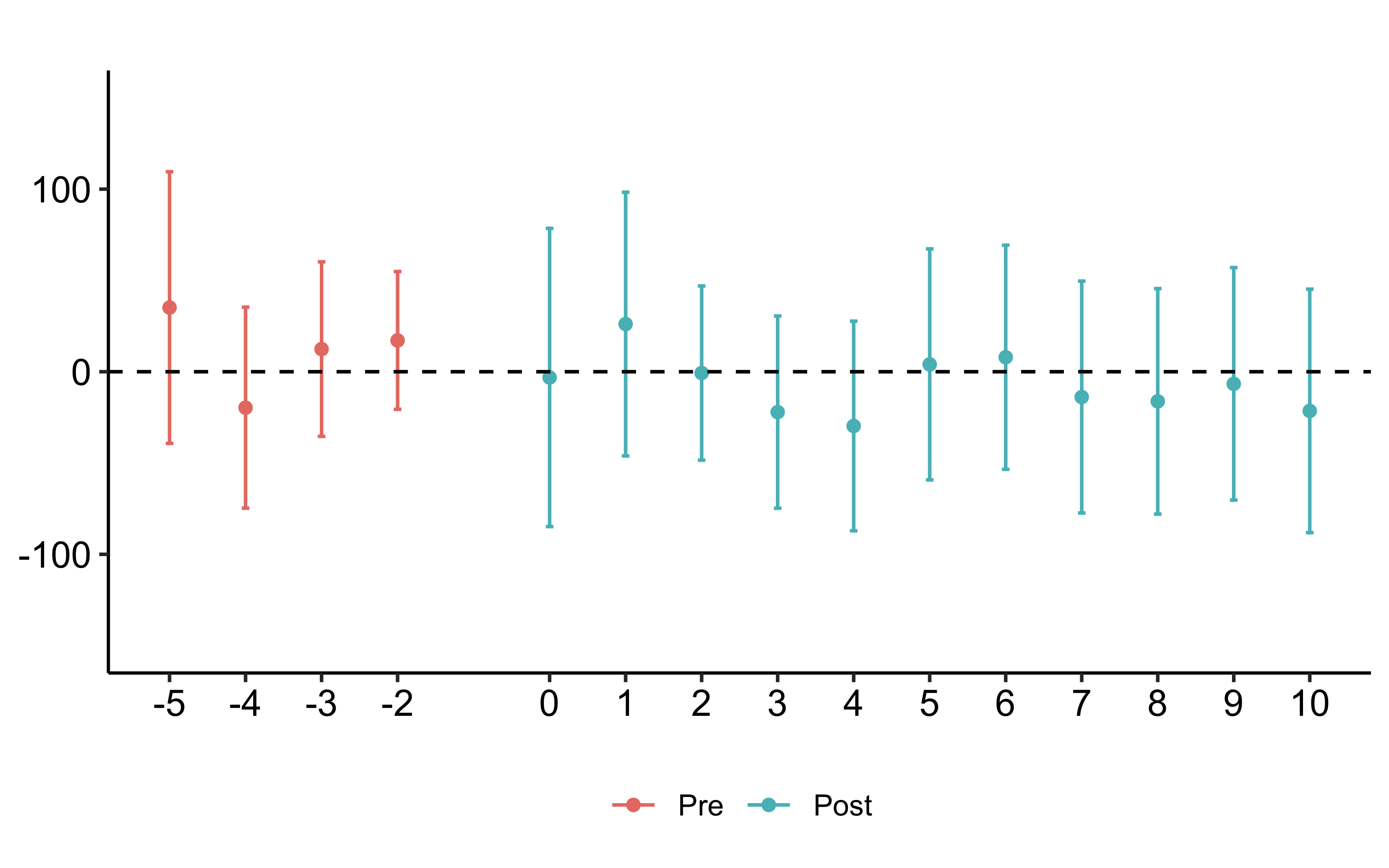}
        \caption*{(b) With Covariates}
    \end{subfigure}

    \label{fig:chained_did_all}
    \floatfoot{\footnotesize{\textbf{Notes}: This figure reports estimates of $\theta_{es}(e)$ for $ATT^{CD}$, $ATT_0^{CD}$, and $ATT_S^{CD}$ obtained using the method described in \hyperref[methods]{Section~\ref{methods}}. Estimates with covariate adjustment are computed using the inverse probability weighting estimator. All parameters are reported with $95\%$ confidence intervals, obtained via multiplier bootstrap with $9{,}999$ replications and clustering at the facility level.}} 
\end{figure}

\hyperref[fig:chained_did_all]{Figure~\ref{fig:chained_did_all}} presents event-study estimates of treatment effects for $ATT^{CD}$, $ATT^{CD}_0$, and $ATT^{CD}_S$. Panel~(a) reports estimates without covariate adjustment, whereas Panel~(b) includes covariates. Specifically, in the latter case, we control for (i) a categorical variable capturing the facility’s main sector of activity among eight broad categories (e.g., animals, energy, mineral industry, etc.), and (ii) a categorical variable indicating the facility's location across four European regions (Western, Eastern, Northern, Southern). Estimates with covariate adjustment are obtained using the inverse probability weighting estimator. All parameters are reported with $95\%$ confidence intervals, obtained via multiplier bootstrap with $9{,}999$ replications, where we cluster at the facility level to account for potential serial correlation in the outcome variable. Further, because the number of treated units is small for some values of $e$, we trim the event study periods that are further from treatment, by restricting $e\in\{-5,- 4,\dots,-2\} \cup \{0, 1, 2,\dots, 10\}$, with the last period before policy implementation as reference period. Across all graphs, the pre-treatment estimates ($e<0$), shown in red, are not statistically different from zero, indicating no evidence of differential trends before the treatment and supporting the parallel trends and limited treatment anticipation assumptions. These results hold irrespective of whether covariates are included, although when conditioning on covariates the confidence bands widen. 
This interpretation is further corroborated by Panel~(b) in \hyperref[average_trajectory]{Figure~\ref*{average_trajectory}}, where the average trajectories of eventually treated and never treated firms appear nearly parallel prior to the introduction of the EU ETS. In the post-treatment period ($e \geq 0$), shown in blue, the estimates of $ATT^{CD}$ are negative and statistically significant when covariates are not included. However, once we condition on covariates, the estimated effect becomes insignificant for most values of $e$. As previously discussed, these estimates are likely to be subject to severe bias due to the presence of spillovers. By contrast, the $ATT_0^{CD}$ estimates are negative and statistically significant in most periods, suggesting a persistent reduction in emissions for firms not subject to spillovers. The effect becomes more pronounced 4-5 years after implementation, pointing at the importance of a long post-treatment horizon. This pattern may be consistent with the gradual adjustments and potential restructuring required to reduce emissions, though identifying the exact mechanisms is beyond the scope of our analysis and would require additional evidence. For $ATT_S^{CD}$, post-treatment estimates are statistically insignificant. The lack of significance for $ATT_S^{CD}$ suggests that the EU ETS had, on average, no impact on reducing emissions among firms engaged in permit trading. Estimates for all three parameters remain virtually unchanged regardless of whether covariates are included. 

\sloppy Results in \hyperref[fig:chained_did_all]{Figure~\ref{fig:chained_did_all}} are complemented by those in \hyperref[overall_att_e]{Table~\ref{overall_att_e}}, which reports the estimates of overall treatment effects based on aggregation schemes of the form $ \theta_{w}^{O} =\frac{1}{\kappa}\sum_{g \in \G}\sum_{t=2}^T\mathbbm{1}\{t\geq g\} \P\left(G_i=g|G\leq T\right)ATT^{CD}(g,t)$, where $\kappa=\sum_{g\in\G}\sum_{t=2}^T\mathbbm{1}\{t\geq g\} \P\left(G_i=g|G\leq T\right)$ for $ATT^{CD}$, $ATT^{CD}_0$, and $ATT^{CD}_S$. That is, $\theta_{w}^{O}$ computes the overall average effect by aggregating all estimated group–time treatment effects across groups and periods for each of the four parameters with weights proportional to group size. The table also reports the estimate of the overall spillover effect, namely $\widehat{AST}^{CD}$. The aggregated value of $\widehat{ATT}^{CD}$, equal to $-48.27$, is statistically different from zero, indicating an overall decrease in total emissions by around $13.5$ per cent in average facility-level emissions when conditioning on covariates. When focusing only on facilities that do not engage into trading activities, the $ATT_0^{CD}$ indicates that the policy led to a statistically significant average reduction by $70.21$ thousand tons of CO\textsubscript{2} per year, which corresponds to $29.3$ per cent decrease in average emissions for this group when conditioning on covariates ($26.4$ per cent of overall emissions). Results confirm a statistically insignificant effect for the overall $ATT_S^{CD}$. The last row of the table reports an estimate of the spillover effect, namely $AST^{CD}$. Interestingly, the coefficient is positive and statistically significant (when not conditioning on covariates). This suggests that the portion of the policy effect attributable to spillovers operates in the direction of increasing emissions, rather than reducing them as one might expect. This finding highlights the importance of accounting for indirect effects when evaluating the overall impact of the policy.

\begin{table}[H]												
\begin{center}												
\caption{\title{Overall Average Treatment Effects} \label{overall_att_e}}												
\begin{tabular}{c*{5}{c}}												
\toprule												
%\vspace{0.05cm}\\												
												
	&	\multicolumn{1}{c}{\textbf{Without covariates}}			&&\multicolumn{1}{c}{\textbf{With Covariates}} &				\textbf{Mean}	& \textbf{N Obs.}		\\
    &			& & &				\textbf{Dep. Var.}	& \\
		\cmidrule(lr){2-2}	\cmidrule(lr){4-4}				\cmidrule(lr){5-5}		\cmidrule(lr){6-6}

\textbf{$\widehat{ATT}^{CD}$}	&	-48.27	***	&	&	-38.17	**	&	283.59	&	7,346	\\
	&	(11.28)		&	&	(15.45)		&		&		\\
												
\vspace{0.05cm}\\												
												
\textbf{$\widehat{ATT}_0^{CD}$}	&	-70.21	***	&	&	-63.23	***	&	239.62	&	6,266	\\
	&	(13.50)		&	&	(16.43)		&		&		\\
												
\vspace{0.05cm}\\												
												
\textbf{$\widehat{ATT}_S^{CD}$}	&	-23.04		&	&	-10.71		&	239.62	&	5,790	\\
	&	(17.75)		&	&	(20.76)		&		&		\\
												
\vspace{0.05cm}\\												
												
\textbf{$\widehat{AST}^{CD}$}	&	26.30	*	&	&	12.70		&	425.83	&	2,006	\\
	&	(14.84)		&	&	(14.42)		&		&		\\
												
%\vspace{0.05cm}\\												
\bottomrule												
												
\end{tabular}												
\floatfoot{\footnotesize{\textbf{Notes}: This table reports estimates of overall treatment effects based on aggregation schemes of the form $\theta_{w}^{O}$ for $ATT^{CD}$, $ATT_0^{CD}$, $ATT_S^{CD}$, and $AST^{CD}$, obtained using the method described in \hyperref[methods]{Section~\ref{methods}}. Covariate-adjusted estimates are computed via the inverse probability weighting estimator. All estimates are reported with $95\%$ confidence intervals, constructed using multiplier bootstrap with $9{,}999$ replications and clustering at the facility level. ***, **, and * denote significance at 1\%, 5\%, and 10\%, respectively.}}												
\end{center}												
\end{table}														
The above results are robust to various changes in modelling choices and sample composition (see Appendix~C). They suggest that, although aggregate CO\textsubscript{2} emissions declined consistent with the effectiveness of the tightening of the emission cap, the trading mechanism did not generate substantial economic incentives for firms to reduce emissions. A plausible explanation is that the cost of acquiring additional permits remained close to zero, thus weakening the incentive for firms exceeding their allocations to undertake abatement efforts.

\subsection{Heterogeneity analysis}
In this subsection we provide the estimates separately for different subsamples to examine heterogeneity in the policy effects. First, we divide the sample into two non-overlapping groups based on whether facilities' average CO\textsubscript{2} emissions during the pre-policy period are above or below the median. Next, using data from the OECD Environmental Policy Stringency (EPS) index, we distinguish between facilities located in countries with relatively stringent pre-existing environmental regulations and those in countries with less stringent regulatory frameworks. Table \ref{overall_att_heterog} reports the  estimates obtained using the covariate-adjusted estimator.
The left panel compares facilities below and above the median level of baseline CO2 emissions. The estimates indicate that the policy effect is concentrated among high-emitting facilities, which experience a clear and persistent decline in emissions in the post-treatment period. By contrast, facilities with emissions below the median do not display a statistically significant change in their own emissions. However, these smaller emitters experience a statistically significant spillover effect. This pattern is consistent with a compliance strategy in which overall adherence to the cap is achieved primarily through emission reductions at high-intensity installations, where abatement opportunities are likely larger and more cost-effective. At the same time, smaller installations appear to comply mainly by purchasing permits rather than by directly reducing emissions, generating spillover effects without a measurable decline in their own emission levels.

The right panel reports analogous results when heterogeneity is defined using the EPS index. In this case, the estimated policy effects are stronger for countries with below-median EPS scores, suggesting that the policy led to larger emission reductions in contexts where pre-existing environmental regulation was relatively weak.
A plausible interpretation is that countries with higher EPS levels had already adopted stricter environmental policies or had achieved lower emissions prior to the introduction of the policy. As a result, these countries faced more limited margins for further adjustment, whereas countries with less stringent regulatory frameworks experienced greater scope for policy-induced reductions.

\begin{table}[H]																														
\begin{center}																														
\caption{\title{Overall Average Treatment Effects -- Heterogeneity Analysis} \label{overall_att_heterog}}																														
\resizebox{\linewidth}{!}{\begin{tabular}{c*{13}{c}}																														
\toprule																														
%\vspace{0.05cm}\\																														
	&	\multicolumn{6}{c}{\textbf{CO\textsubscript{2} Emissions}}													&	& 	\multicolumn{6}{c}{\textbf{Stringency Index}}													\\
		\cmidrule(lr){2-7}															\cmidrule(lr){9-14}													
	&	\multicolumn{3}{c}{\textbf{Below Median}}						&\multicolumn{3}{c}{\textbf{Above Median}} 							&	&	\multicolumn{3}{c}{\textbf{Below Median}}							&\multicolumn{3}{c}{\textbf{Above Median}} 						\\
		\cmidrule(lr){2-4}						\cmidrule(lr){5-7}									\cmidrule(lr){9-11}							\cmidrule(lr){12-14}						
	&	$\theta_{w}^{O}$		&	\textbf{Mean}	&	\textbf{N Obs.}	&	$\theta_{w}^{O}$		&	\textbf{Mean}	&	\textbf{N Obs.}	&	&	$\theta_{w}^{O}$		&	\textbf{Mean}	&	\textbf{N Obs.}	&	$\theta_{w}^{O}$		&	\textbf{Mean}	&	\textbf{N Obs.}	\\
	&			&	\textbf{Dep. Var.}	&		&			&	\textbf{Dep. Var.}	&		&	&			&	\textbf{Dep. Var.}	&		&			&	\textbf{Dep. Var.}	&		\\
    \cmidrule(lr){2-4}						\cmidrule(lr){5-7}									\cmidrule(lr){9-11}							\cmidrule(lr){12-14}	
\textbf{$\widehat{ATT}_0^{CD}$}	&	-4.16		&	28.79	&	4520	&	-77.94	***	&	537.44	&	2507	&	&	-75.72	***	&	221.24	&	3102	&	19.33		&	270.64	&	2810	\\
	&	(3.47)		&		&		&	(27.02)		&		&		&	&	(20.02)		&		&		&	(27.88)		&		&		\\
\vspace{0.01cm}\\																														
\textbf{$\widehat{ATT}_S^{CD}$}	&	-0.91		&	28.95	&	4217	&	-1.70		&	550.44	&	2313	&	&	-48.86		&	205.84	&	2735	&	28.04		&	298.03	&	2709	\\
	&	(3.05)		&		&		&	(28.74)		&		&		&	&	(26.87)		&		&		&	(35.95)		&		&		\\
\vspace{0.01cm}\\																														
\textbf{$\widehat{AST}^{CD}$}	&	9.76	**	&	70.25	&	960	&	10.75		&	550.85	&	1480	&	&	38.86	*	&	392.12	&	1076	&	-31.19		&	459.90	&	866	\\
	&	(4.08)		&		&		&	(16.19)		&		&		&	&	(23.40)		&		&		&	(19.56)		&		&		\\
\bottomrule																														
\end{tabular}}																														
\floatfoot{\footnotesize{\textbf{Notes}: This table reports estimates of overall treatment effects based on aggregation schemes of the form $\theta_{w}^{O}$ for $ATT^{CD}$, $ATT_0^{CD}$, $ATT_S^{CD}$, and $AST^{CD}$, obtained using the method described in \hyperref[methods]{Section~\ref{methods}}. Covariate-adjusted estimates are computed via the inverse probability weighting estimator. All estimates are reported with $95\%$ confidence intervals, constructed using multiplier bootstrap with $9{,}999$ replications and clustering at the facility level. ***, **, and * denote significance at 1\%, 5\%, and 10\%, respectively.}}																														
\end{center}																														
\end{table}																														
\section{Conclusions}\label{conclusions}

This paper has developed and implemented a DiD framework that explicitly allows for spillovers among treated firms to separate direct policy effects from indirect trading-induced effects, and validated the proposed estimator in Monte Carlo exercises. The empirical application shows that the EU ETS generated meaningful emission reductions only among firms that did not engage in permit trading (the direct effect), while once spillovers from allowance trading are accounted for the average effect of the policy across treated firms is close to zero. This pattern suggests that the trading mechanism attenuated the cap’s incentive to reduce emissions in practice. Heterogeneity analysis further indicates that that the policy effect is stronger for large emitters and for facilities located in countries with below-median EPS index, suggesting that plants facing weaker pre-existing regulations had more room to adjust and achieve emission reductions.
These results have important consequences for policy makers. If allowance trading substantially weakens incentives to reduce emissions for many regulated firms, policy makers should reconsider the design of the policy to ensure that the trading mechanism does not systematically offset the cap's environmental goals. Finally, the paper highlights the need to account for strategic interactions among regulated entities when designing and evaluating cap-and-trade policies.
\newpage

\begin{spacing}{1.0}

\bibliographystyle{apalike}

\bibliography{Bibliography}

\end{spacing}

\clearpage
\appendix
\label{Appendix}
\input{Appendix}
\newpage
\input{Appendix_B}

\newpage
\input{Appendix_C}

\end{document}

%% file: Appendix.tex
\section*{Appendix A -- Proofs of the main results}\label{Appendix_econometrics}
\renewcommand{\theequation}{A.\arabic{equation}}

%----------------------------------------------------------------------------------------
%	PROOF THEOREM 1
%----------------------------------------------------------------------------------------
\renewcommand{\thesubsection}{A.\arabic{subsection}}
\subsection{Proofs of Theorem 1}\label{Appendix_proof_Th_1}
\begin{proof}[Proof of Theorem 1]
This proof follows directly from the proof of Theorem 1 in \cite{Bellegoetal2025}, with minor adaptations. 
Under \hyperref[as_RS]{Assumption~\ref*{as_RS}}, the data-generating process consists of draws from the mixture distribution $F_M(\cdot)$ defined as

\begin{equation*}
    \sum_{t=2}^T \lambda_{t-1,t} F_{Y_{t-1},Y_{t}, G_2,\dots,G_T, C, S, X\mid A_{it-1,t}=1}\left(y_t,y_{t+1}, g_2,\dots,g_T, c, s, x \mid A_{t-1,t}=1\right)
\end{equation*}
where $\lambda_{t-1,t}=\P(A_{t-1,t}=1)$ is the probability of being sampled in $t-1$ and $t$. Under the above DGP, expectations under the mixture distribution do not generally correspond to population expectations. This discrepancy arises due to time-varying $\lambda_{t-1,t}$ and because \hyperref[as_Missing_trends]{Assumption~\ref*{as_Missing_trends}} does not impose independence between unobserved heterogeneity in $\yit$ and the sampling process. In what follows, we denote by $\E_M\left[\cdot\right]$ the expectation taken with respect to the mixture distribution $F_M\left[\cdot\right]$. Now, define $ATT_{0,\X}(g,\tau) = \E\left[Y_{i\tau}(g,\0) - Y_{i\tau}(0,\0) \mid \X_i, G_{ig} = 1, \Si = 0 \right]$, and consider its first-difference representation as
    \begin{equation*}
        \begin{aligned}
            \Delta ATT_{0,\X}(g, \tau) =&  ATT_{0,\X}(g, \tau)-ATT_{0,\X}(g, \tau-1) \\
                                  =& \E\left[\yitau(g,\0)-\yitau(0,\0)\mid \X_i, G_{ig}=1, \Si=0\right]-\\
                                  &\E\left[Y_{i\tau-1}(g,\0)-Y_{i\tau-1}(0,\0)\mid \X_i, G_{ig}=1, \Si=0\right] \\
                                  =& \E\left[\yitau(g,\0)-Y_{i\tau-1}(g,\0)\mid \X_i, G_{ig}=1, \Si=0\right]-\\
                                  &\E\left[\yitau(0,\0)-Y_{i\tau-1}(0,\0)\mid \X_i, G_{ig}=1, \Si=0\right]\\
                                  =& \E\left[\yitau(g,\0)-Y_{i\tau-1}(g,\0)\mid \X_i, G_{ig}=1 , \Si=0 \right]-\\
                                  &\E\left[\yitau(0,\0)-Y_{i\tau-1}(0,\0)\mid \X_i, C_{i}=1, \Si=0\right]\\
                                  =& \E\left[\yitau-Y_{i\tau-1}\mid \X_i, G_{ig}=1, \Si=0, A_{i\tau-1,\tau}=1\right]-\\
                                  &\E\left[\yitau-Y_{i\tau-1}\mid \X_i, C_{i}=1,\Si=0, A_{i\tau-1,\tau}=1\right]\\
                                  =& A_{\X}(g,\tau)-B_{\X}(g, \tau),
        \end{aligned}
    \end{equation*}

\noindent where the third equality results from a simple rearrangement of terms; the fourth follows from \hyperref[as_PT]{Assumption~\ref*{as_PT}} holding; and the fifth equality follows from Assumptions~\hyperref[as_RS]{\ref*{as_RS}}-\hyperref[as_NA]{\ref*{as_NA}}.

We can use the above expression to develop $ATT_0(g,t)$ into

\begin{equation}
    \begin{aligned}
        ATT_0(g,t) & = \E\left(Y_{it}\left(g,\mathbf{0}\right)-Y_{it}\left(0,\mathbf{0}\right)\mid G_{ig}=1, \Si = 0 \right)\\
                   & = \E\left(\E\left[Y_{it}\left(g,\mathbf{0}\right)-Y_{it}\left(0,\mathbf{0}\right)\mid \X_i, G_{ig}=1, \Si = 0 \right] \mid  G_{ig}=1, \Si = 0\right)\\
                   & = \E\left(ATT_{0,\X}(g,t)\mid  G_{ig}=1, \Si = 0\right)\\
                   &=\E\left(\sum_{\tau=g}^t \Delta ATT_{0,\X}(g,\tau)\mid  G_{ig}=1, \Si = 0\right) \\
                   &=\sum_{\tau=g}^t \E\left(\Delta ATT_{0,\X}(g,\tau)\mid  G_{ig}=1, \Si = 0, A_{i\tau-1,\tau}=1\right)\\
                   &=\sum_{\tau=g}^t \E\left(A_{\X}(g,\tau)-B_{\X}(g, \tau)\mid  G_{ig}=1, \Si = 0, A_{i\tau-1,\tau}=1\right),
    \end{aligned}
\end{equation}
with 

\begin{equation}
    \begin{aligned}\label{WeightA}
        \E\left (A_{\X}(g,\tau)\mid  G_{ig}=1, \Si = 0, A_{i\tau-1,\tau}=1\right) & = \E\left[\yitau-Y_{i\tau-1}\mid G_{ig}=1, \Si=0, A_{i\tau-1,\tau}=1\right] \\
        & = \E_M\left[\frac{G_{ig}(1-\Si) A_{i\tau-1,\tau}}{\E\left[G_{ig}(1-\Si) A_{i\tau-1,\tau}\right]}\left(\yitau-Y_{i\tau-1}\right)\right]
    \end{aligned}    
\end{equation}

where the equivalence in \hyperref[WeightA]{(\ref{WeightA})} follows from the law of iterated expectations, the definition of $F_M$, and the fact that for any random variable $Z$, and dummy variables $F$ and $T$, the conditional expectation with respect to the event $\{F=0,T=1\}$ can be written as $\E(Z|F=0,T=1)=\frac{\E((1-F)Z|T=1)}{\E((1-F)|T=1)}$. Suppressing, for notational convenience, the conditioning on $= 1$ (e.g., $G_{ig} = 1$), defining $\Delta Y_{\tau}:=\yitau-Y_{i\tau-1}$, the second term can be developed as follows

\begin{equation}
    \begin{aligned}\label{WeightB}
        &\E\left(B_{\X}(g, \tau)\mid  G_{ig}=1, \Si = 0, A_{i\tau-1,\tau}=1\right)  =\E\left(\E\left[\Delta Y_{\tau}\mid \X_i, C_i,\Si=0, A_{i\tau-1,\tau}\right]\mid  G_{ig}, \Si = 0, A_{i\tau-1,\tau}\right) \\
        &=\E\left(\E\left[\frac{C_i}{\E\left(C_i\mid \X_i,\Si=0, A_{i\tau-1,\tau}\right)}\Delta Y_{\tau}\bigl\vert \X_i,\Si=0, A_{i\tau-1,\tau}\right]\bigl\vert  G_{ig}, \Si = 0, A_{i\tau-1,\tau}\right) 
        \\ &=\frac{\E\left(\Gig\E\left[\frac{C_i}{\E\left(C_i\mid \X_i,\Si=0, A_{i\tau-1,\tau}\right)}\Delta Y_{\tau}\bigl\vert \X_i,\Si=0, A_{i\tau-1,\tau}\right]\Bigl\vert  \Si = 0, A_{i\tau-1,\tau}\right)} {\E\left(\Gig \mid \Si=0, A_{i\tau-1,\tau}\right)} \\     
        &=\frac{\E\left(\E\left(\Gig\mid \X_i, \Si = 0, A_{i\tau-1,\tau}\right)\E\left[\frac{C_i}{\E\left(C_i\mid \X_i,\Si=0, A_{i\tau-1,\tau}\right)}\Delta Y_{\tau}\bigl\vert \X_i,\Si=0, A_{i\tau-1,\tau}\right]\Bigl\vert  \Si = 0, A_{i\tau-1,\tau}\right)}{\E\left(\Gig \mid \Si=0, A_{i\tau-1,\tau}\right)} \\ 
        &=\frac{\E\left(\E\left[\E\left(\Gig\mid \X_i, \Si = 0, A_{i\tau-1,\tau}\right)\frac{C_i}{\E\left(C_i\mid \X_i,\Si=0, A_{i\tau-1,\tau}\right)}\Delta Y_{\tau}\bigl\vert \X_i,\Si=0, A_{i\tau-1,\tau}\right]\Bigl\vert  \Si = 0, A_{i\tau-1,\tau}\right)}{\E\left(\Gig \mid \Si=0, A_{i\tau-1,\tau}\right)} \\ 
        &=\frac{\E\left(\E\left(\Gig\mid \X_i, \Si = 0, A_{i\tau-1,\tau}\right)\frac{C_i}{\E\left(C_i\mid \X_i,\Si=0, A_{i\tau-1,\tau}\right)}\Delta Y_{\tau}\Bigl\vert  \Si = 0, A_{i\tau-1,\tau}\right)}{\E\left(\Gig \mid \Si=0, A_{i\tau-1,\tau}\right)}.
    \end{aligned}    
\end{equation}
Using the definition of generalized propensity score $\P_g^{s}(\X)$ and the fact that $\Gig+C_i=1$

\begin{equation}
    \begin{aligned}
        \dots &=\frac{\E\left(\frac{C_i\E\left(\Gig\mid \X_i, \Gig+C_i, \Si = 0, A_{i\tau-1,\tau}\right)}{\E\left(C_i\mid \X_i,  \Gig+C_i, \Si=0, A_{i\tau-1,\tau}\right)}\Delta Y_{\tau}\Bigl\vert  \Si = 0, A_{i\tau-1,\tau}\right)}{\E\left(\Gig \mid \Si=0, A_{i\tau-1,\tau}\right)} \\ 
        &=\frac{\E\left(\frac{C_i\P_g^{0}(\X)}{1-\P_g^{0}(\X)}\Delta Y_{\tau}\Bigl\vert  \Si = 0, A_{i\tau-1,\tau}\right)}{\E\left(\Gig \mid \Si=0, A_{i\tau-1,\tau}\right)}\\
        &=\frac{\E\left(\frac{(1-\Si)C_i\P_g^{0}(\X)}{1-\P_g^{0}(\X)}\Delta Y_{\tau}\Bigl\vert  A_{i\tau-1,\tau}\right)}{\E\left(\Gig \mid \Si=0, A_{i\tau-1,\tau}\right)\E\left(1-\Si\mid A_{i\tau-1,\tau}\right)}\\
        &=\frac{\E\left(\frac{(1-\Si)A_{i\tau-1,\tau}C_i\P_g^{0}(\X)}{1-\P_g^{0}(\X)}\Delta Y_{\tau}\right)}{\E\left(\Gig \mid \Si=0, A_{i\tau-1,\tau}\right)\E\left(1-\Si\mid A_{i\tau-1,\tau}\right)\E(A_{i\tau-1,\tau})}\\
        &=\E_M\left(\frac{C_i(1-\Si)A_{i\tau-1,\tau}}{\E_M\left(\Gig(1-\Si)A_{i\tau-1,\tau}\right) }\frac{\P_g^{0}(\X)}{1-\P_g^{0}(\X)}\Delta Y_{\tau}\right),
    \end{aligned}    
\end{equation}
where 
\begin{equation*}
\begin{aligned}
    &\E_M\left[\frac{C_i(1-\Si)A_{i\tau-1,\tau}\P_g^{0}(\X)}{1-\P_g^{0}(\X)}\right] =\frac{\E_M\left(C_i(1-\Si)A_{i\tau-1,\tau}\E_M\left(\Gig\mid \X_i, \Gig+C_i, \Si = 0, A_{i\tau-1,\tau}\right)\right)}{\E_M\left(C_i\mid \X_i, \Gig+C_i, \Si = 0, A_{i\tau-1,\tau}\right)} \\
    &= \frac{\E_M\left[C_i(1-\Si)A_{i\tau-1,\tau}\E_M\left(\Gig (1-\Si) \mid \X_i, A_{i\tau-1,\tau}\right)\right] }{\E_M\left(C_i\mid \X_i, \Si = 0, A_{i\tau-1,\tau}\right)\E_M\left(1-\Si\mid \X_i, A_{i\tau-1,\tau}\right)}\\
    &= \frac{\E_M\left[C_i(1-\Si)A_{i\tau-1,\tau}\E_M\left(\Gig (1-\Si) A_{i\tau-1,\tau} \mid \X_i\right)\right]}{\E_M\left(C_i\mid \X_i, \Si = 0, A_{i\tau-1,\tau}\right)\E_M\left(1-\Si\mid \X_i, A_{i\tau-1,\tau}\right)\E_M(A_{i\tau-1,\tau}\mid \X_i)} \\
    &= \frac{\E_M\left[\E_M(C_i(1-\Si)A_{i\tau-1,\tau}\mid \X_i)\E_M\left(\Gig (1-\Si) A_{i\tau-1,\tau} \mid \X_i\right)\right]}{\E_M\left(C_i\mid \X_i, \Si = 0, A_{i\tau-1,\tau}\right)\E_M\left(1-\Si\mid \X_i, A_{i\tau-1,\tau}\right)\E_M(A_{i\tau-1,\tau}\mid \X_i)} \\
     &= \E_M\left(\frac{\E_M(C_i(1-\Si)A_{i\tau-1,\tau}\mid \X_i)\E_M\left(\Gig (1-\Si) A_{i\tau-1,\tau} \mid \X_i\right)}{\frac{\E_M\left(C_i(1-\Si)A_{i\tau-1,\tau}\mid \X_i\right)}{\E_M(1-\Si\mid \X_i, A_{i\tau-1,\tau})\E_M(A_{i\tau-1,\tau}\mid \X_i)}\E_M\left(1-\Si\mid \X_i, A_{i\tau-1,\tau}\right)\E_M(A_{i\tau-1,\tau}\mid \X_i)}\right) \\
     &= \E_M\left(\E_M\left(\Gig (1-\Si) A_{i\tau-1,\tau} \mid \X_i\right)\right) \\
     &= \E_M\left(\Gig (1-\Si) A_{i\tau-1,\tau}\right). \\
\end{aligned}
\end{equation*}
\noindent This completes the proof.
    
\end{proof}

%----------------------------------------------------------------------------------------
%	GENERAL MISSING DATA PATTERNS
%----------------------------------------------------------------------------------------
\subsubsection*{General missing data patterns}

In some settings, consecutive observations may not be available to the researcher. In such cases, identification and inference based on one-period differences are infeasible. Nevertheless, the framework presented in the main text can be extended to accommodate more general missing data patterns, allowing for units observed at distant time periods and for situations where covariates are thought to be related with outcome dynamics and are distributed differently across groups. Specifically, following \cite{Bellegoetal2025}, we can define the $k$-period differences, denoted by $\Delta_k ATT_0(g, t)$, which represent the long difference between periods $t - k$ and $t$, for $k = 1, \dots, t - 1$. In this more flexible setting, we no longer require that units be observed in two consecutive periods; instead, we only require that each unit be observed in at least two (not necessarily adjacent) periods. 

Under this framework, each parameter $ATT_0(g, t)$ can be identified in multiple ways using available $k$-period differences, for $1 \leq k \leq t - 1$. This creates an overidentified linear inverse problem, which--when addressed correctly--can lead to efficiency gains by leveraging all available information across subsamples. 

To formalise this, let  $\Delta \mathbf{ ATT}_0$ be the vector stacking all estimable $\Delta_k ATT_0(g, t)$, for all $t \geq 2$ and $1 \leq k \leq t - 1$, and let $L_\Delta$ be its length. By construction, each $\Delta_k ATT_0(g, t)$ satisfies $\Delta_k ATT_0(g, t)=ATT_0(g,t)-ATT_0(g,t-k)$. So the system can be written in matrix form as
\begin{equation}
\Delta\mathbf{ ATT}_0=\mathbf{WATT}_0    
\end{equation}
where $\mathbf{ATT}_0$ is the vector of the $ATT_0(g, t)$ parameters of interest, of length $L \leq L_\Delta$, and $\mathbf{W}$ is a known matrix with elements in $\{-1, 0, 1\}$.

To solve this overidentified system, \cite{Bellegoetal2025} propose a GMM approach. Let $\Omega$ denote the covariance matrix of $\Delta\mathbf{ATT}_0$; then the optimal GMM estimator of $\mathbf{ATT}_0$ is given by $\widehat{\mathbf{ATT}_0}=\left(\mathbf{W}'\hat{\mathbf{\Omega}}^{-1}\mathbf{W}\right)^{-1}\mathbf{W}'\hat{\mathbf{\Omega}}^{-1} \widehat{\Delta\mathbf{ATT}_0}$. We consider two versions of this estimator: one using the identity matrix as the weighting matrix, and the efficient version using the estimator of $\mathbf{\Omega}$ as proposed in \cite{Bellegoetal2025}.

The proposed method allows identify $ATT_0(g,t)$ in  more general setting where covariates are allowed to be correlated with  sample selection. However, we must modify \hyperref[as_Missing_trends]{Assumption~\ref*{as_Missing_trends}} as follows

\begin{assumption}[\textit{Missing Trends at Random 2}]\label{as_Missing_trends2}
For all $t=1,\dots, T-1$ and $k=1,\dots,t-1$,
\begin{equation}
    A_{it-k,t} \indep Y_{it}-Y_{it-k} \mid \mathbf{X}_i, \Gig, \Si.
\end{equation}
\end{assumption}

%----------------------------------------------------------------------------------------
%	ATTS & AST
%----------------------------------------------------------------------------------------
\subsection{Identification of \texorpdfstring{$ATT^{CD}_S$}{ATT CDS} and  \texorpdfstring{$AST^{CD}$}{AST CD}}
In this section, we extend the identification results presented in \hyperref[methods]{Section~\ref{methods}} to $ATT^{CD}_S$ and $AST^{CD}_S$. 
While identification of $ATT^{CD}_S$ relies on the same assumptions required for $ATT^{CD}_0$, identification of $AST^{CD}_S$ requires a modification of the conditional parallel trends assumption, as stated below

\begin{assumption}[\textit{Modified Conditional Parallel Trends}]\label{as_PT2} Let $\delta$ be as defined in \hyperref[as_NA]{Assumption~\ref*{as_NA}}. For each $g\in \G$ and $t \in \{2,\dots, T\}$ such that $ t\geq g-\delta$,
\begin{equation}
     \E\left( Y_{it} \left(g,\0 \right)- Y_{it-1}\left(0,\0 \right)\mid  \X_i,   G_{ig}=1, \Si=1\right)=\E\left(Y_{it}\left(g,\0 \right)- Y_{it-1}\left(0,\0 \right)\mid  \X_i,  \Gig=1, \Si=0 \right) \quad \text{a.s.}
\end{equation}
\end{assumption}

\hyperref[as_PT2]{Assumption~\ref*{as_PT2}} requires that, conditional on $\mathbf{X}_i$ and cohort membership $G_{ig}=1$, the average change in the treated potential outcome without interference is the same for units that are eventually exposed to spillovers 
and for units that are not exposed to spillovers.

\subsubsection*{Identification of $ATT^{CD}_S$}
Define the following weights

$$w_{i\tau-1,\tau}^G(g)= \frac{G_{ig}A_{i\tau-1,\tau}\Si}{\E_M\left[G_{ig}A_{i\tau-1,\tau}\Si\right]},$$

\noindent and

$$w_{i\tau-1,\tau}^C(g, \X)= \frac{\P_g^{1}(\X)C_i\Si A_{i\tau-1,\tau}}{1-\P_g^{1}(\X)}\Big/\E_M\left[\frac{\P_g^{1}(\X)C_i\Si A_{i\tau-1,\tau}}{1-\P_g^{1}(\X)}\right].$$

\begin{theorem}\label{Th2}
Under Assumptions~\hyperref[as_irreversibility]{\ref*{as_irreversibility}}--\hyperref[as_Overlap]{\ref*{as_Overlap}}, and for $g\in\G$ and $t\in\{2,\dots, T\}$, the long-term average treatment effect with spillovers in period $t$ is nonparametrically identified and given by
\begin{equation}
    ATT_S^{CD}(g,t)=\sum_{\tau=g-\delta}^t \Delta ATT_S(g,\tau),
\end{equation}

\sloppy \noindent
where $\Delta ATT_S(g,t)=\E_M\left[w_{i\tau-1,\tau}^G\left(g\right)\left(Y_{i\tau}-Y_{i\tau-1}\right)\right]-\E_M\left[w_{i\tau-1,\tau}^C\left(g, \X\right)\left(Y_{i\tau}-Y_{i\tau-1}\right)\right]$. 
\end{theorem}

\hyperref[Th2]{Theorem~\ref*{Th2}} mirrors \hyperref[Th1]{Theorem~\ref*{Th1}} and establishes that $ATT^{CD}_S$ is identified under the same set of assumptions required for the identification of $ATT^{CD}_0$.

\begin{proof}[Proof of Theorem 2]
This proof follows directly from the proof of \hyperref[Th1]{Theorem~\ref*{Th1}} provided above, with the only difference being that the conditioning is now on $\Si = 1$ rather than on $\Si = 0$.
\end{proof}

\subsubsection*{Identification of $AST^{CD}$}
Define the following weights

$$w_{i\tau-1,\tau}^{G,1}(g)= \frac{G_{ig}A_{i\tau-1,\tau}\Si}{\E_M\left[G_{ig}A_{i\tau-1,\tau}\Si\right]},$$

\noindent and

$$w_{i\tau-1,\tau}^{G,0}(g)= \frac{\Gig(1-\Si) A_{i\tau-1,\tau}}{\E_M\left[\Gig(1-\Si)A_{i\tau-1,\tau}\right]}.$$

\begin{theorem}\label{Th3}
Under Assumptions~\hyperref[as_irreversibility]{\ref*{as_irreversibility}}--\hyperref[as_NA]{\ref*{as_NA}} and \hyperref[as_PT2]{\ref*{as_PT2}}, and for $g\in\G$ and $t\in\{2,\dots, T\}$, the long-term average spillover effect on the treated in period $t$ is nonparametrically identified and given by
\begin{equation}
    AST^{CD}(g,t)=\sum_{\tau=g-\delta}^t \Delta AST(g,\tau),
\end{equation}

\sloppy \noindent
where $\Delta AST(g,t)=\E_M\left[w_{i\tau-1,\tau}^{G,1}\left(g\right)\left(Y_{i\tau}-Y_{i\tau-1}\right)\right]-\E_M\left[w_{i\tau-1,\tau}^{G, 0}\left(g\right)\left(Y_{i\tau}-Y_{i\tau-1}\right)\right]$. 
\end{theorem}

\begin{proof}[Proof of Theorem 3]
This proof follows directly from the proof of \hyperref[Th1]{Theorem~\ref*{Th1}} provided above, with only minor adaptations.
\end{proof}

%% file: Appendix_B.tex
\section*{Appendix B}\label{Appendix_B}
\renewcommand{\theassumption}{B.\arabic{assumption}}
\setcounter{assumption}{0}
\renewcommand{\thetable}{B.\arabic{table}}
\setcounter{table}{0}
\setcounter{subsection}{0}
% Subsection numbering
\renewcommand{\thesubsection}{B.\arabic{subsection}}

\subsection{Construction of the CO2 Emission Variable}

Information on emissions comes from the E-PRTR database. We restrict our analysis to facilities that emit greenhouse gases, which are recognized under the \cite{protocol1997kyoto} as contributors to global warming. The database provides facility-level information on annual greenhouse gas emissions (in kilograms). Each facility may emit different types of greenhouse gases in a given year (e.g. carbon dioxide (CO\textsubscript{2}), hydrofluorocarbons (HFCs), etc.). However, these raw data are not directly comparable between gases. A standard approach is to express emissions in terms of carbon dioxide equivalents (CO\textsubscript{2}-eq), which converts the emissions of each greenhouse gas into the amount of CO\textsubscript{2} that would produce the same global warming potential (GWP). This allows for comparability across gases and aggregation into a common metric of total emissions. Carbon dioxide equivalents are typically reported in million metric tonnes (MMTCDE).\footnote{For further details, see \url{https://ec.europa.eu/eurostat/statistics-explained/index.php?title=Glossary:Carbon_dioxide_equivalent}.} Formally, the CO\textsubscript{2} equivalent of a gas is obtained by multiplying its emissions (in tonnes) by the associated GWP:

\begin{equation*}
    \text{MMTCDE} = (\text{million metric tonnes of a gas}) \cdot (\text{GWP of the gas}).
\end{equation*}

For example, the 100-year GWP of methane is 21. This implies that emissions of 1 million metric tonnes of methane correspond to emissions of 21 million metric tonnes of carbon dioxide equivalents.
To convert greenhouse gas emissions from the E-PRTR database, we first transform kilograms into tonnes by multiplying the quantity reported for each facility and year by $0.001$. We then apply the conversion factors for the global warming potential (GWP) provided in the Global Warming Potentials (IPCC Second Assessment Report) of the United Nations Climate Change.\footnote{Interested readers can refer to \url{https://unfccc.int/process/transparency-and-reporting/greenhouse-gas-data/greenhouse-gas-data-unfccc/global-warming-potentials}.}
However, for hydrofluorocarbons (HFCs) and perfluorocarbons (PFCs), the E-PRTR does not report the specific chemical (e.g., perfluoromethane or perfluoroethane) but only the group classification. To address this, we compute three alternative CO\textsubscript{2}-equivalent measures: one using the minimum, one using the median, and one using the maximum value of the 100-year GWP for the corresponding group.
Throughout the paper, we rely on the conversion based on the median 100-year GWP values for each group. We ignore the other two conversions (minimum and maximum) since the pairwise correlations across the three measures exceed $0.99$, making them virtually indistinguishable.

%% file: Appendix_C.tex
\section*{Appendix C}\label{Appendix_C}
\renewcommand{\theassumption}{C.\arabic{assumption}}
\setcounter{assumption}{0}
\renewcommand{\thetable}{C.\arabic{table}}
\setcounter{table}{0}

\subsection*{C.1 Additional tables}
\renewcommand{\theassumption}{C.\arabic{assumption}}
\setcounter{assumption}{0}
\renewcommand{\thetable}{C.\arabic{table}}
\setcounter{table}{0}
\renewcommand{\thefigure}{C.\arabic{figure}}
\setcounter{figure}{0}

\begin{table}[H]
\begin{center}
\caption{\title{Descriptive Statistics on Staggered Treatment Adoption} \label{staggered_tr_adop}}
\begin{tabular}{c*{2}{c}}
\toprule
  \textbf{Year of Treatment}          &       \textbf{Frequency}&     \textbf{Percent}\\
\cline{1-3}
%Never       &        3590&       62.09\\
2005        &        1625&       22.12\\
2006        &         427&        5.81\\
2007        &          76&        1.03\\
2008        &         131&        1.78\\
2009        &          52&        0.71\\
2010        &          32&        0.44\\
2011        &          27&        0.37\\
2012        &          29&        0.39\\
2013        &         162&        2.21\\
2014        &          36&        0.49\\
2015        &           5&        0.07\\
2016        &           2&        0.03\\
2017        &           3&        0.04\\
2018        &          10&        0.14\\
2019        &          17&        0.23\\
2020        &           2&        0.03\\
%\cline{1-3}
%Total       &        7346&     100    \\
\bottomrule
\end{tabular}
\end{center}
\end{table}

\begin{table}[htbp]
\begin{center}
\caption{\title{Descriptive Statistics on Leavers} \label{leavers}}
\begin{tabular}{c*{2}{c}}
\toprule
    \textbf{Year of Exit}          &\textbf{Frequency}&\textbf{Percent}\\
\cline{1-3}
2007        &           3&        1.49\\
2008        &           4&        1.98\\
2009        &          32&       15.84\\
2010        &           8&        3.96\\
2011        &           2&        0.99\\
2012        &           4&        1.98\\
2013        &          33&       16.34\\
2014        &          69&       34.16\\
2015        &          30&       14.85\\
2016        &          17&        8.42\\
\cline{1-3}
Total       &         202&      100.00\\
\bottomrule
\end{tabular}
\end{center}
\end{table}

\subsection*{C.2 Robustness checks}
\renewcommand{\theassumption}{C.\arabic{assumption}}
\setcounter{assumption}{0}
\renewcommand{\thetable}{C.\arabic{table}}
\setcounter{table}{0}

\begin{figure}[H]
    \centering
    \caption{Chained DiD Estimates -- Robustness Checks}
    
    % Row 1 title
   % \vspace{0.2cm}
    {\small (i) Never Treated \par\vspace{-0.05cm}}
    \begin{subfigure}{0.48\linewidth}
        \centering
        \includegraphics[width=0.8\textwidth]{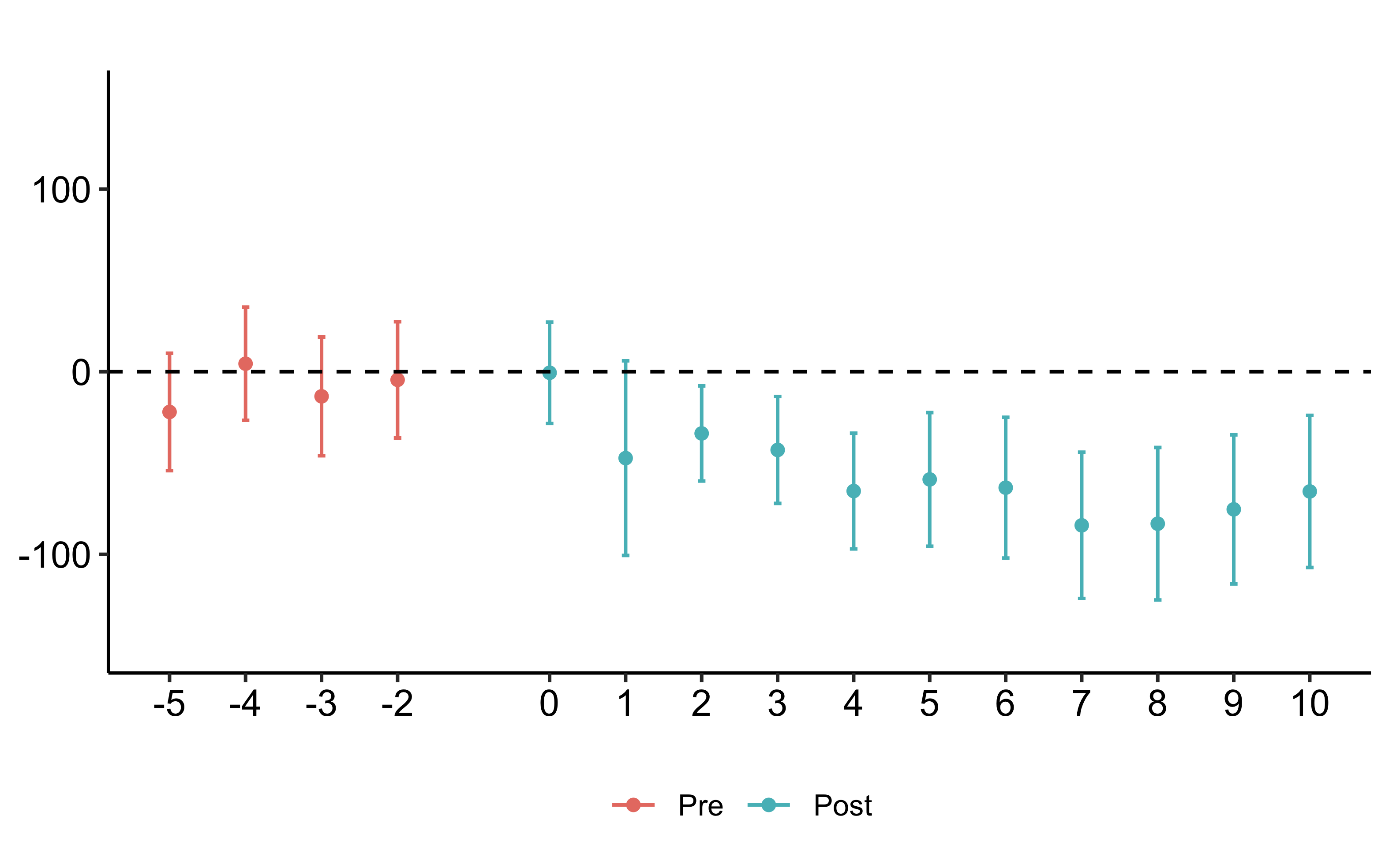}
        \caption*{(a) $\widehat{ATT}_0^{CD}$}
    \end{subfigure}
    \hfill
    \begin{subfigure}{0.48\linewidth}
        \centering
        \includegraphics[width=0.8\textwidth]{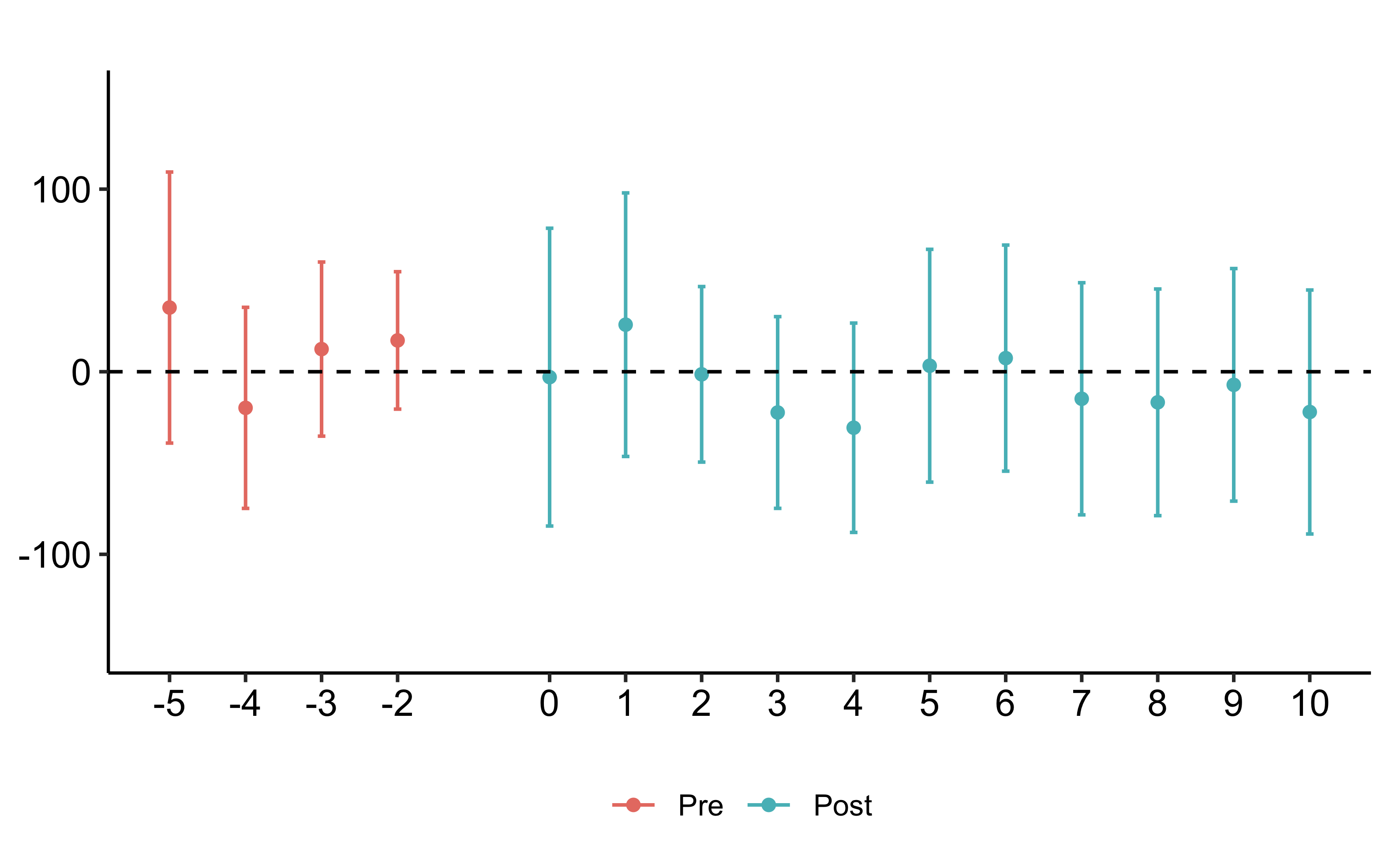}
        \caption*{(b) $\widehat{ATT}_S^{CD}$}
    \end{subfigure}

   % \vspace{0.2cm}

    % Row 2 title
     {\small (ii) Exact calendar year of entry \par\vspace{-0.05cm}}
    \begin{subfigure}{0.48\linewidth}
        \centering
        \includegraphics[width=0.8\textwidth]{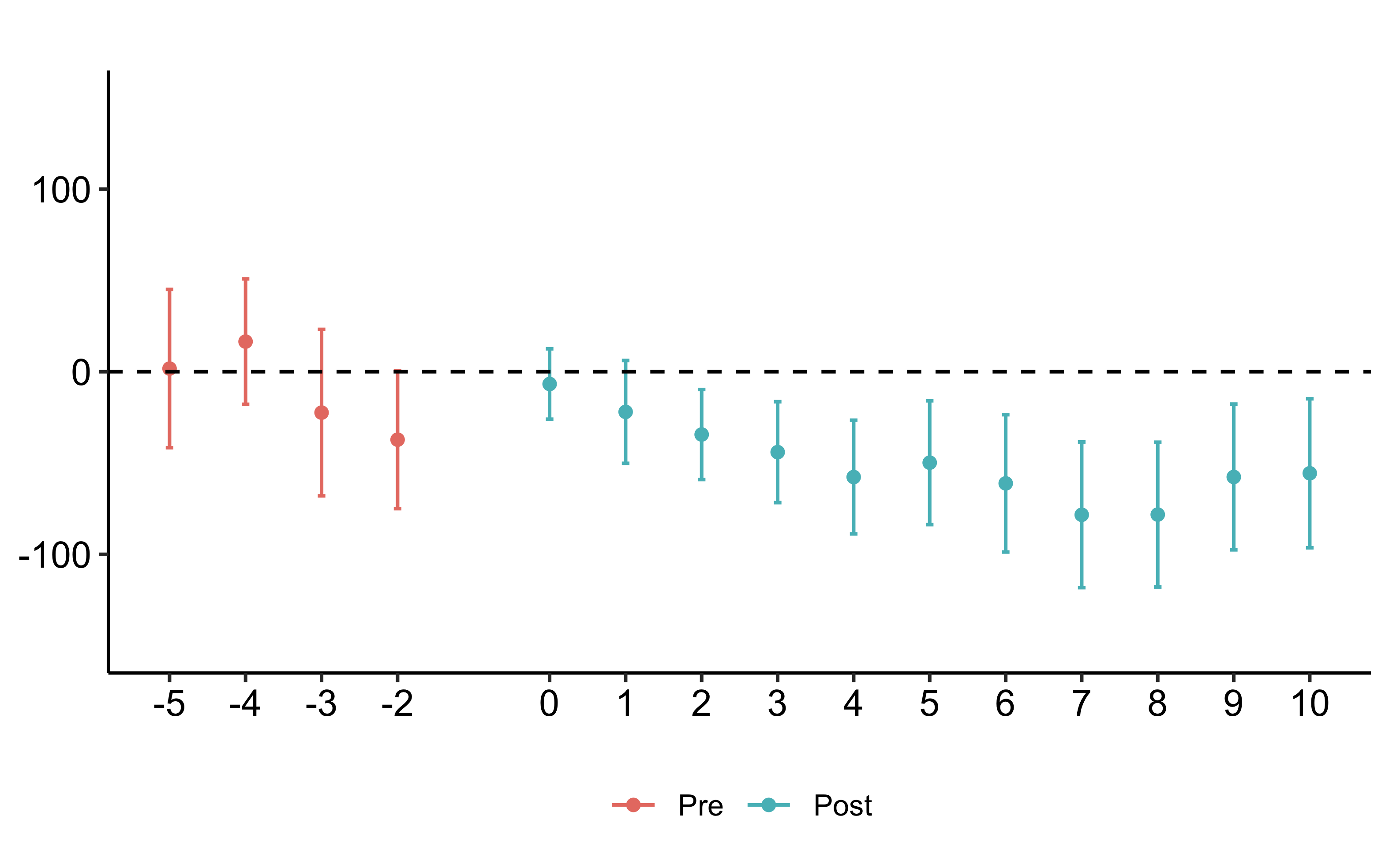}
        \caption*{(a) $\widehat{ATT}_0^{CD}$}
    \end{subfigure}
    \hfill
    \begin{subfigure}{0.48\linewidth}
        \centering
        \includegraphics[width=0.8\textwidth]{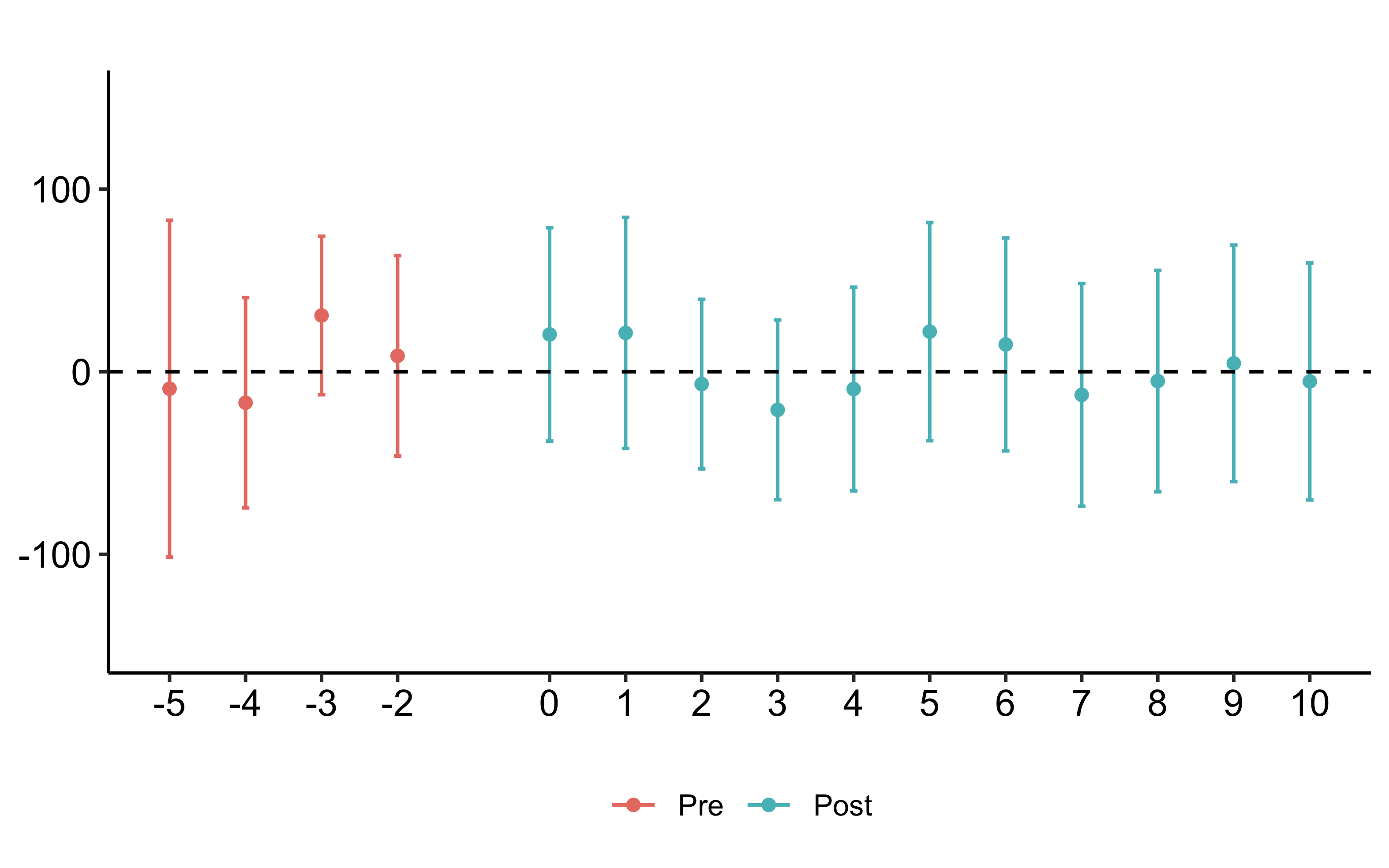}
        \caption*{(b) $\widehat{ATT}_S^{CD}$}
    \end{subfigure}

 %   \vspace{0.2cm}
    
    % Row 3 title
    {\small (iii) $\theta_{es}^{bal}(e,e')$ \par\vspace{-0.05cm}}
    \begin{subfigure}{0.48\linewidth}
        \centering
        \includegraphics[width=0.8\textwidth]{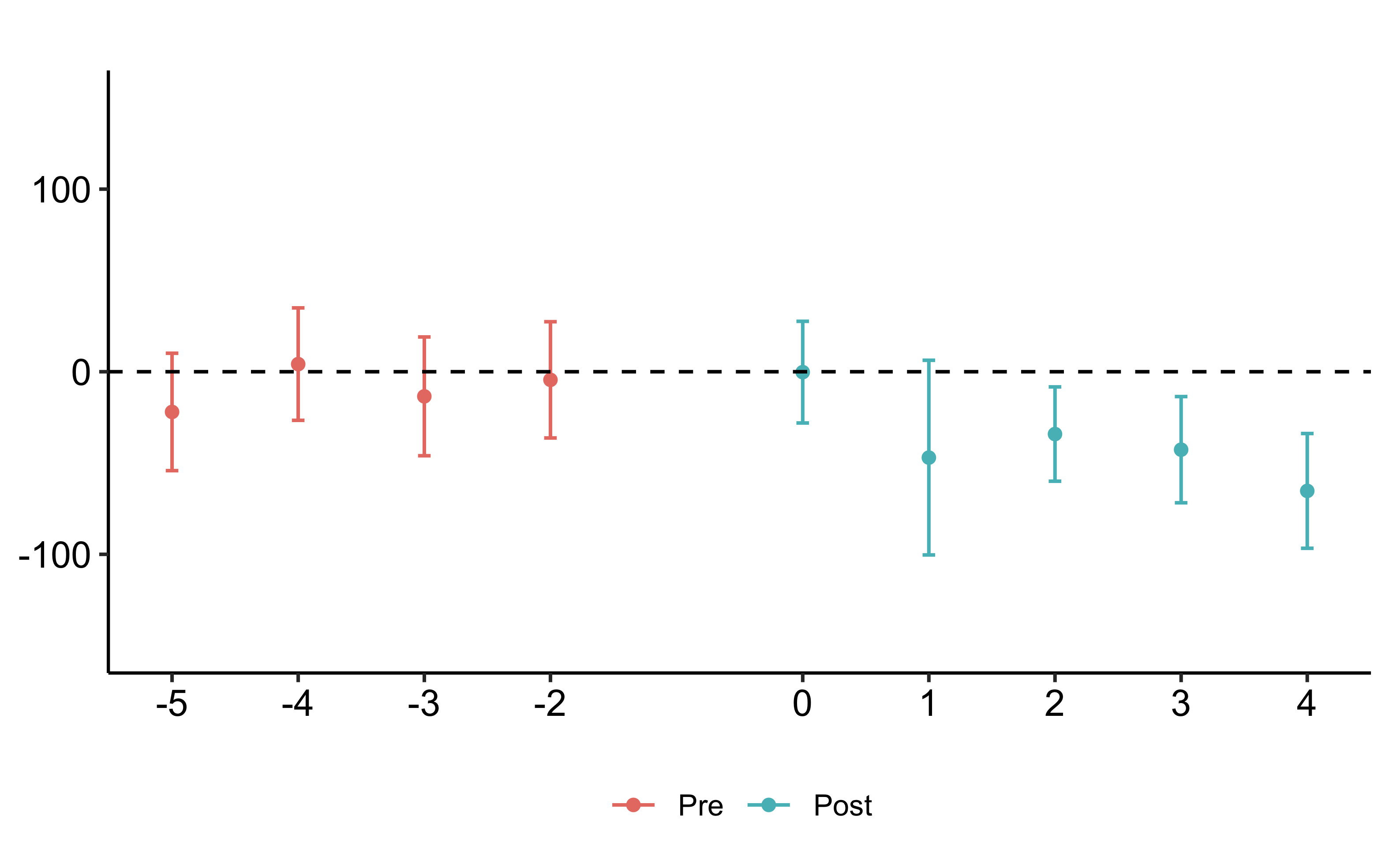}
        \caption*{(a) $\widehat{ATT}_0^{CD}$}
    \end{subfigure}
    \hfill
    \begin{subfigure}{0.48\linewidth}
        \centering
        \includegraphics[width=0.8\textwidth]{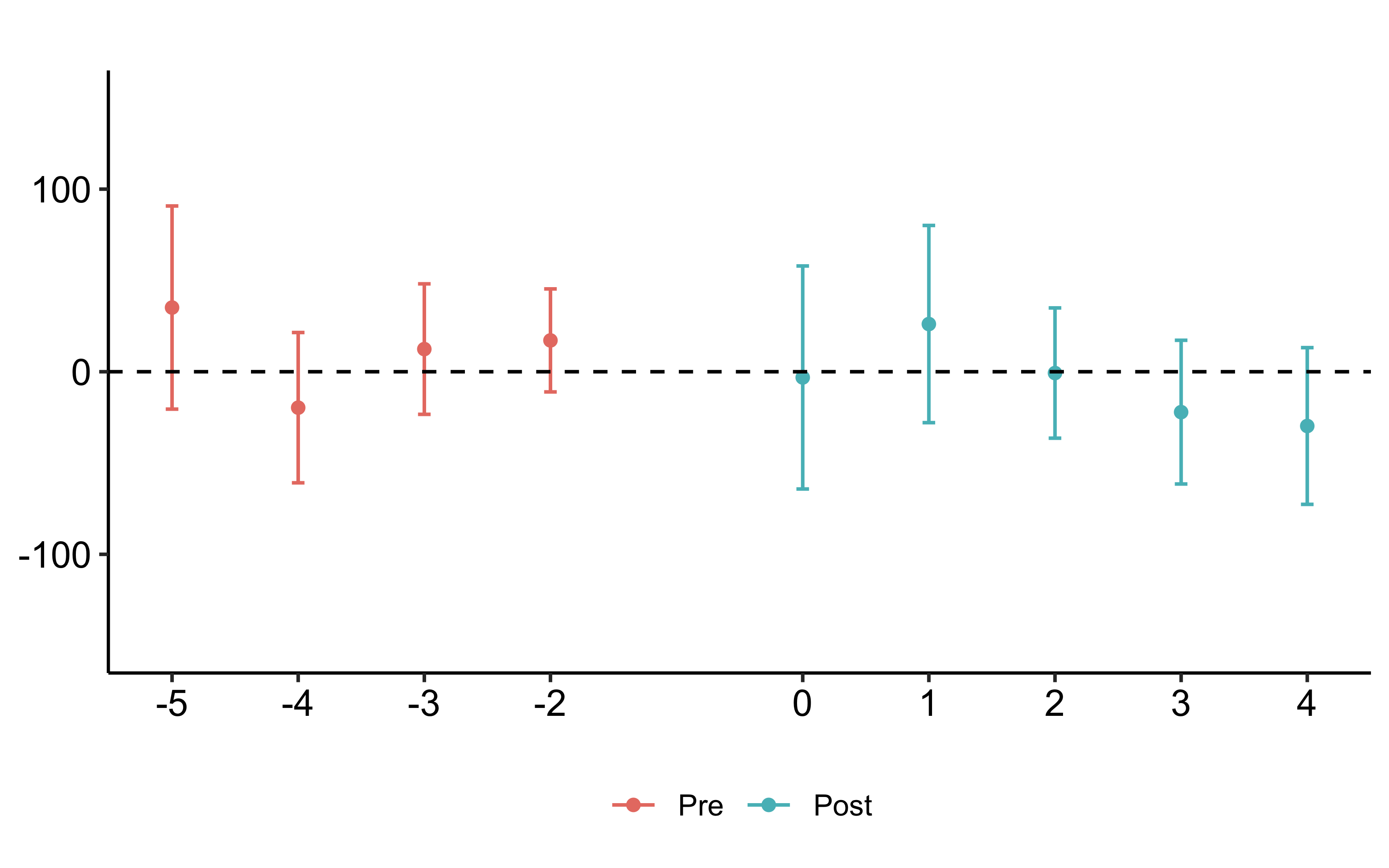}
        \caption*{(b) $\widehat{ATT}_S^{CD}$}
    \end{subfigure}

  %  \vspace{0.2cm}
    
    % Row 4 title
    {\small (iv) Excluding Leavers \par\vspace{-0.05cm}}

    \begin{subfigure}{0.48\linewidth}
        \centering
        \includegraphics[width=0.8\textwidth]{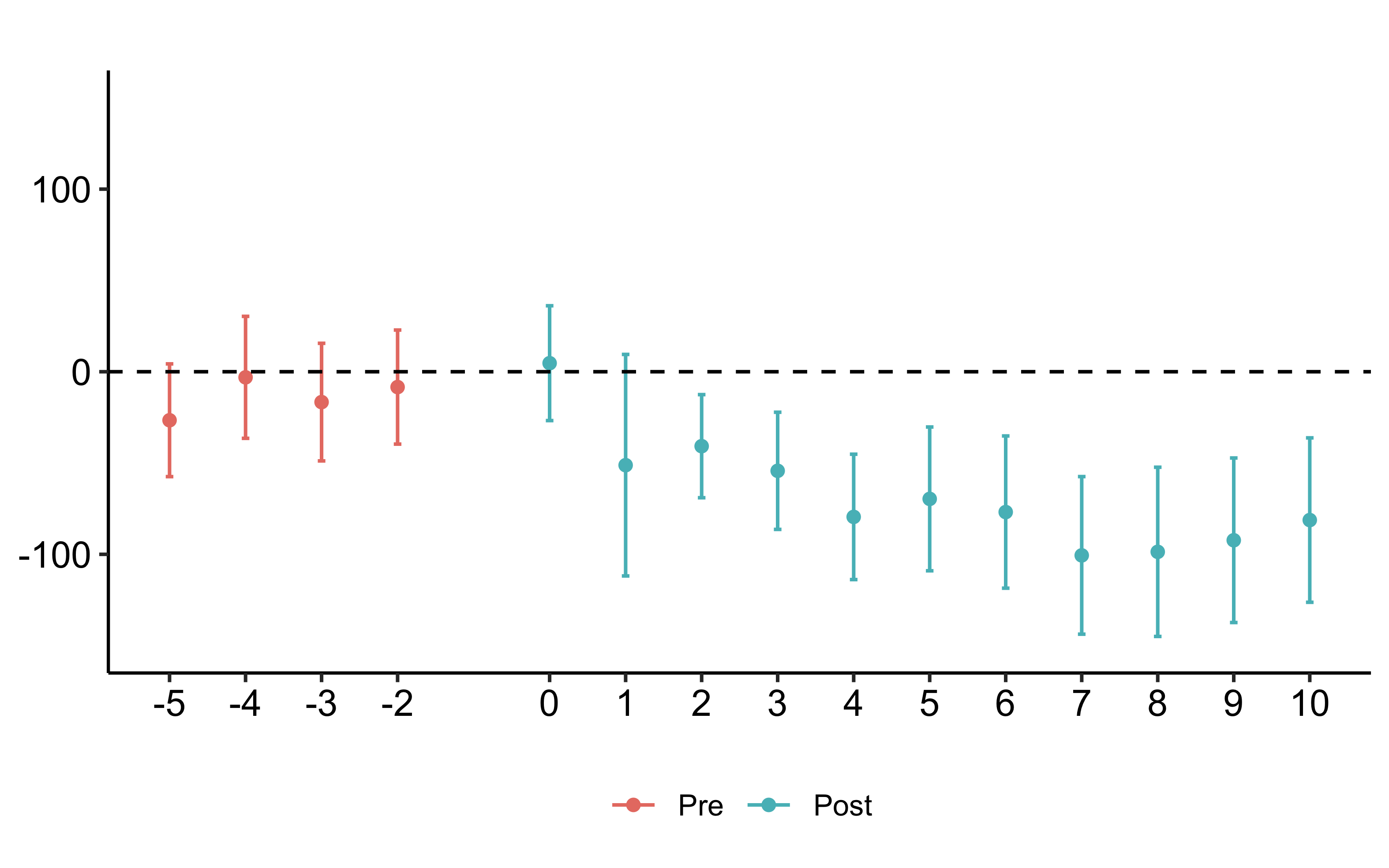}
        \caption*{(a) $\widehat{ATT}_0^{CD}$}
    \end{subfigure}
    \hfill
    \begin{subfigure}{0.48\linewidth}
        \centering
        \includegraphics[width=0.8\textwidth]{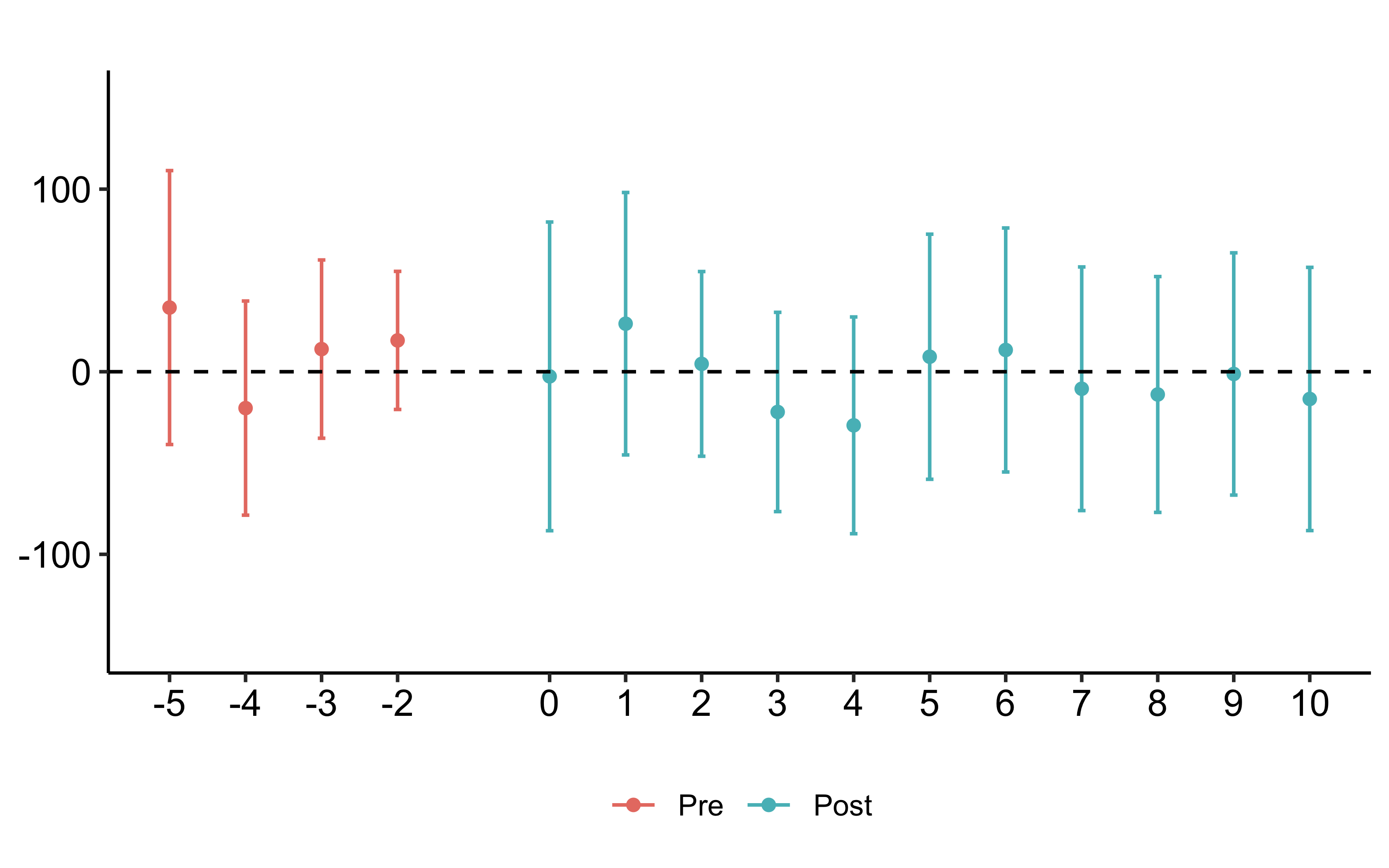}
        \caption*{(b) $\widehat{ATT}_S^{CD}$}
    \end{subfigure}

   % \vspace{0.2cm}

    \label{fig:chained_did_robchecks}
\floatfoot{\footnotesize \textbf{Notes}: 
%\arraystretch{0.3}
This figure presents event-study estimates of $ATT_0^{CD}$ (panel (a)) and $ATT_S^{CD}$ (panel (b)), obtained using the method described in \hyperref[methods]{Section~\ref{methods}} with covariate adjustment. Covariate adjustment is implemented via the inverse probability weighting estimator. All estimates are reported with $95\%$ confidence intervals, computed using a multiplier bootstrap with $9{,}999$ replications and clustering at the facility level.}
\end{figure}

We now carry out a small robustness analysis to check that our main results are not driven by modelling choices or sample composition. Panels~(i)–(iv) in \hyperref[fig:chained_did_robchecks]{Figure~\ref{fig:chained_did_robchecks}} present the results of four sensitivity exercises for the $ATT_0^{CD}$ (Column~(a)) , and $ATT_S^{CD}$ estimates (Column~(b)). In all scenarios, we focus on estimates obtained with covariate adjustment, as results without covariates show similar patterns. 
Further, we omit the results for $ATT^{CD}$, as they display the same qualitative pattern as those presented above. Specifically, the $ATT^{CD}$ estimates are negative and statistically significant (only in some cases), but smaller in magnitude than the $ATT_0^{CD}$, reflecting the severe bias arising from spillovers. 

In \hyperref[fig:chained_did_all]{Figure~\ref{fig:chained_did_all}}, we report the results obtained using the not-yet-treated as the benchmark group. While relying on not-yet-treated units expands the pool of valid comparison units--potentially improving inference due to their likely similarity in pre-treatment trends with already-treated units--this approach has an important drawback. If units that are treated later adjust their behaviour in response to the policy’s introduction, the never-treated group may provide a more reliable counterfactual. One reason that led us to use EU ETS phases rather than exact calendar time was precisely to account, at least partially, for potential anticipatory behaviour among firms that entered later within the same phase. Nevertheless, it may still be the case that later-treated units, even if to a lesser extent, anticipate their future participation in the scheme. For this reason, we re-run the analysis using never-treated units as the benchmark group. Results are shown in Panel~(i) of \hyperref[fig:chained_did_robchecks]{Figure~\ref{fig:chained_did_robchecks}}. We find that the estimates are almost unchanged relative to those in \hyperref[fig:chained_did_all]{Figure~\ref{fig:chained_did_all}}, suggesting that, if anticipatory behaviours are present, they represent only a minor concern.

In Panel~(ii) of \hyperref[fig:chained_did_robchecks]{Figure~\ref{fig:chained_did_robchecks}}, we present the results obtained using the exact calendar year of entry rather than defining the starting year of treatment based on EU ETS phases. Although the group-time average treatment effects remain broadly in line with those reported in \hyperref[fig:chained_did_all]{Figure~\ref{fig:chained_did_all}}, when considering firms not subject to spillovers, we now observe that, for some values of $e<0$, the estimates are statistically different from zero  at the $10\%$. This pattern may arise for two reasons. First, it may reflect the relatively small number of newly treated firms in certain calendar years, which causes some pre-treatment periods to be imprecisely estimated. Second, it may indicate a violation of the Parallel Trends assumption, with differential pre-treatment trends arising from anticipatory behaviour of later-treated firms within the same EU ETS phase. This finding corroborates our initial choice to define the treatment start date according to the EU ETS phase in which a firm joins the scheme.

One limitation of using aggregation schemes of the form $\theta_{es}(e)$ is that, when plotting $\theta_{es}(e)$ across different values of $e$, compositional changes may occur,  potentially complicating the interpretation of these parameters. Such changes arise because the set of groups contributing to the estimate at each event time may differ \citep[see][for a detailed discussion]{Callaway2021}. For instance, at later event times, only groups treated early can still be observed, which may cause the estimates to reflect changes in the sample composition rather than purely the effect of treatment duration. To address this issue, we \emph{balance} the groups with respect to $e$, i.e., we compute group-time average treatment effects over a fixed set of groups that remain in the sample throughout the entire event-study horizon. In Panel~(iii) of \hyperref[fig:chained_did_robchecks]{Figure~\ref{fig:chained_did_robchecks}}, we implement this balancing procedure by retaining only those groups exposed to treatment for at least four periods. Specifically, we report event-study estimates based on aggregation schemes of the form $\theta_{es}^{bal}(e,e') =\sum_{g\in\G}\mathbbm{1}\{g+e'\leq T\} \P\left(G=g|G+e'\leq T\right)ATT_0^{CD}(g,g+e)$ with $0\leq e \leq e' \leq T-2$. Hence, $\theta_{es}^{bal}(e,e')$ computes the $ATT_0^{CD}$ for firms with event time $e$ that remain in treatment for at least $e'$ periods. The same applies to $ATT_S^{CD}$. The results in panel (iii) are fully consistent with those in \hyperref[fig:chained_did_all]{Figure~\ref{fig:chained_did_all}}. This suggests that our findings are not driven by changes in group composition across different values of $e$.

Up to this point, we have assumed that treatment is an absorbing state. However, as shown in \hyperref[leavers]{Table~\ref*{leavers}}, some firms exit treatment during the post-treatment period. To account for this, in Panel~(iv) of \hyperref[fig:chained_did_robchecks]{Figure~\ref{fig:chained_did_robchecks}} we replicate the analysis after excluding these firms. Although the parameters are estimated with less precision--largely due to the exclusion of 196 facilities--the results remain fully consistent with those in \hyperref[fig:chained_did_all]{Figure~\ref{fig:chained_did_all}}. The robustness of our findings, even after excluding leavers, supports our initial hypothesis that once a facility is treated, its behaviour is permanently affected.

A similar reasoning applies to spillovers. While we have assumed that spillovers are irreversible, in practice about $72.42\%$ of treated facilities in our sample stopped trading for at least one period after they began. To account for this, we replicate the analysis excluding these facilities as well. Again, the results remain robust and consistent with the notion that, as with treatment assignment, once firms are affected by spillovers, their behaviour is permanently altered. 
Lastly, since our CO\textsubscript{2} data end in 2017, we excluded from the main analysis all facilities that appear in our data set but only enter treatment after 2021 (i.e., during the fourth EU ETS phase). As an additional robustness, we re-run the analysis including this subset of facilities. The results remain consistent with those presented in \hyperref[fig:chained_did_all]{Figure~\ref{fig:chained_did_all}}. Additional results are not reported but are available upon request.

